\documentclass[12pt,twoside]{article}
\usepackage{epsf,epsfig,rotating}
\begin{document}

%%%%%%%%%%%%%%%%%%%%%%%%%%%%%%%%%%%%%%%%%%%%%%%%%%%%%%%%%%%%%%%%%%%%%%
\newcommand{\lsim}   {\mathrel{\mathop{\kern 0pt \rlap
  {\raise.2ex\hbox{$<$}}}
  \lower.9ex\hbox{\kern-.190em $\sim$}}}
\newcommand{\gsim}   {\mathrel{\mathop{\kern 0pt \rlap
  {\raise.2ex\hbox{$>$}}}
  \lower.9ex\hbox{\kern-.190em $\sim$}}}
\def\be{\begin{equation}}
\def\ee{\end{equation}}
\def\ba{\begin{eqnarray}}
\def\ea{\end{eqnarray}}
\def\d{{\rm d}}
\def\ap{\approx}

\font\menge=bbold9 scaled \magstep1
\def\nota#1{\hbox{$#1\textfont1=\menge $}}
\def\P{\nota P}

%%%%%%%%%%%%%%%%%%%%%%%%%%%%%%%%%%%%%%%%%%%%%%%%%%%%%%%%%%%%%%%%%%%%%%%%%
\title{\hfill {\small CERN-TH 2001-239}\\
       \vskip1.0cm
       Extensive Air Showers from Ultra High Energy Gluinos}
\author{V. Berezinsky$^{1}$, M. Kachelrie{\ss}$^{2}$, 
        and S. Ostapchenko$^{3,4}$ \\
        {\it\small $^1$INFN, Lab. Naz. del Gran Sasso, I--67010
          Assergi (AQ)} \\
        {\it\small $^2$TH Division, CERN, CH--1211 Geneva 23} \\
        {\it\small $^3$Forschungszentrum Karlsruhe, 
          Institut f\"ur Kernphysik, D--76021 Karlsruhe, Germany}  \\
        {\it\small $^4$Moscow State University, Institute of Nuclear Physics,
        199899 Moscow, Russia} \\
        }

\date{August 31, 2001}
\maketitle
\begin{abstract}
We study the proposal that the cosmic ray primaries above the
Greisen-Zatsepin-Kuzmin (GZK) cutoff are gluino-containing hadrons 
($\tilde g$-hadrons). We describe the interaction of $\tilde
g$-hadrons with nucleons in the framework of the Gribov-Regge approach 
using a modified version of the hadronic interaction model QGSJET for
the generations of Extensive Air Showers (EAS). There are two mass
windows marginally allowed for gluinos: $m_{\tilde g}\lsim 3$~GeV 
and $25\lsim m_{\tilde g}\lsim 35$~GeV. Gluino-containing hadrons 
corresponding to the second window produce EAS very different from
the observed ones. Light $\tilde g$-hadrons corresponding to the 
first gluino window produce EAS similar to those initiated 
by protons, and only future detectors can marginally distinguish
them. We propose a beam-dump accelerator experiment to search for 
$\tilde g$-hadrons in this mass window. We emphasize the importance 
of this experiment: it can discover (or exclude) the light gluino and 
its role as a cosmic ray primary at ultra high energies.
\end{abstract}

%Cosmic-rays, Other particles (including hypothetical) 

\newpage
\section{Introduction}
Since long time light gluinos have attracted attention as a possible carrier 
of the very high energy signal in the universe. In the 80s, they were
studied as a possible primary particle from Cyg X-3 \cite{Aur85,BI86}, now  
as a primary particle of the observed Ultra High Energy Cosmic Rays (UHECR)
\cite{Ch98,Fa98}. 

The observations of UHECR with energies above $10^{20}$~eV impose a
serious problem (see \cite{reviews} for recent reviews).

The data show the 
presence of a new, nearly isotropic component in the UHECR flux above
the energy $E\sim 10^{19} $~eV \cite{reviews}. Since the
arrival directions of the UHECR show no correlation with the galactic
plane and the galactic magnetic field cannot isotropize particles of
such energies, this component is thought to be extragalactic. 
On the other hand, the signature of extragalactic protons, the 
Greisen-Zatsepin-Kuzmin (GZK) cutoff~\cite{GZK} at 
$E\simeq 6\cdot 10^{19}$~eV, is not found. The other natural UHE
primaries, nuclei and photons, must also suffer a similar
cutoff. Meanwhile, four different UHECR experiments \cite{reviews}
do not show the presence of such a cutoff. The two highest energy events are 
detected by  AGASA \cite{AGASA94} and  Fly's Eye \cite{FE95} at energy 
$2\cdot 10^{20}~$eV and $3\cdot 10^{20}~$eV, respectively. The total
number of events with energy higher  than $1\cdot 10^{20}~$eV is about
30, 17 of which are detected by AGASA \cite{AGASA01}. The accuracy of
the energy 
determination is estimated to be better than 20--30\%. The energies
of the two highest energy events \cite{AGASA94} and \cite{FE95} are
determined very reliably.

To resolve this puzzle, it seems that 
new ideas in astrophysics or particle physics are required.

The proposals involving particle physics include UHE particles from
superheavy dark matter~\cite{SDM} and topological defects~\cite{TD},
the resonant interaction of UHE neutrinos with dark matter
neutrinos~\cite{rnu}, strongly interacting neutrinos~\cite{snu}, 
new particles as UHE primaries~\cite{Ch98,Fa98,newp} and such a
radical possibility as Lorentz invariance violation~\cite{Loren}. 
(For more references see also the reviews cited in \cite{reviews}.) 

In this paper, we shall consider a  gluino-containing hadron
($\tilde{g}$-hadron) as a carrier of the cosmic UHE signal, being inspired
by the correlation  between AGN and arrival directions  of UHE
particles suggested by the analyses in Refs.~\cite{Fa98,corr}. This
correlation implies that the signal carrier is neutral and is not
absorbed by the CMBR. The light gluino is a suitable candidate for
such a primary: it can be efficiently produced in  pp-interactions in
astrophysical sources, it is not strongly absorbed by CMBR (see below)
and it produces EAS in the atmosphere very similar to those observed. 

Heavy gluinos are naturally produced in decays of superheavy
particles~\cite{Be98}. 

We shall study here the interaction of both light and heavy gluinos
with nucleons at UHE.  
In most interesting applications gluinos must be light (see below). 
To be a suitable primary of UHECR, the $\tilde{g}$-hadron should
satisfy three conditions:

\begin{enumerate}
\item The longitudinal shower profile of Fly's Eye highest energy event with
  $E=3\cdot 10^{20}$~eV is well fitted by a proton
  \cite{halz95,kalm95}, though  Fly's Eye collaboration does not exclude 
  a photon as a primary \cite{FE95}.
  Therefore, $\tilde{g}$-hadrons should
  essentially mimic proton (or photon) induced air showers. 
\item 
 To shift the GZK cutoff to higher energies, the new hadron should
 have a mass in excess of the proton mass: 
 the threshold energy for any energy-loss reaction on microwave photons 
 increases with increasing primary mass, while the fraction of energy
 lost per scattering decreases. Moreover, it is desirable that its
 cross-section for interactions with CMBR photons is smaller
 than the proton's one.  This can be achieved if, e.g., the mass of the
 first resonance $X$ that can be excited in the reaction $\tilde
 g$-hadron$+\gamma_{\rm CMBR}\to X$ is relatively large.
\item
 The primary has to be stable or quasi-stable with lifetime 
 $\tau\gsim 10^6 {\rm s} \: (m/{\rm GeV})\: (L/{\rm Gpc})$ in order to
 survive its travel from a source (e.g. AGN) at distance 
 $L\sim 100$--1000~Mpc to the Earth.
\end{enumerate}

In principle gluino-containing hadrons ($\tilde{g}$-hadrons)
could satisfy the above requirements. Below we shall
shortly review the status of $\tilde{g}$-hadrons as UHECR signal
carrier.

To satisfy the third condition, the gluino should be the Lightest
Supersymmetric Particle (LSP), or have a very small mass difference
with the LSP. 
It also can be the second lightest supersymmetric particle, if 
the LSP is the gravitino; in this case the gluino decays gravitationally
and its lifetime can be long enough. 
Theoretically the best motivated candidates for the LSP are the
neutralino and gravitino. While in minimal
supergravity models the LSP is the lightest neutralino (in some
part of the parameter space it is the sneutrino), in models with
gauge-mediated SUSY the LSP is normally the gravitino.
In Farrar's model~\cite{Fa96}, the gluino is the LSP because the 
dimension-three SUSY breaking terms are set to zero. 
A theoretically more appealing scenario containing a light gluino was
developed in Refs.~\cite{Ra98,Ra99}. There, the gluino with mass
1--100~GeV was found in a SO(10) model with gauge-mediated SUSY
breaking and Higgs-messenger mixing. In this model either the gluino
or the gravitino is the LSP. In the latter case, the gluino can decay
but has a sufficiently long lifetime to be a viable UHECR primary, 
$\tau\sim 100$~yr. 

In a physical state, the gluino is bound into colourless hadrons. What is
the lightest state of gluino-containing hadrons? 

In the 80s (see \cite{BI86}), it was argued on the basis of QCD sum rules
that the {\em glueballino} $\tilde{g}g$ is the lightest  $\tilde{g}$-hadron.
The lightest baryonic state, {\em gluebarino}, was demonstrated to be
the $\bar{g}uud$-hadron  
\cite{VO}. Gluebarino is a long-lived particle because its decay 
needs either violation of baryon number or R-parity \cite{VO}.    
More recently, Farrar
proposed \cite{Fa96} the neutral hadron $S^0$, a $\tilde{g}uds$ bound-state,
as the lightest $\tilde{g}$-hadron (see also the calculations in the MIT
bag model of Ref.~\cite{Bu85}).  

There is some controversy if a light gluino, with a mass of a few GeV, is
allowed. As it stands, the Farrar model \cite{Fa96} is in conflict with
searches for glueballino decays \cite{Ad97,Fa99,Al99} as well as for 
decays of other
unstable $\tilde g$-hadrons \cite{Al97}. However, these searches 
had been restricted to a narrow band of lifetimes and masses, and their
results are not valid  in 
the context of more generic models.

The existence of a light gluino ($m_{\tilde g}\lsim 5$~GeV)
can be (dis-) proved due to its contribution to the running of
$\alpha_s$ and to QCD colour coefficients in a practically
model-independent way. 
The authors of Ref.~\cite{Cs97} used the ratio $R$ between the
hadronic and the $\bar\mu\mu$ production cross-section in $e^+e^-$
annihilation at different energies to constrain the light gluino
scenario. They excluded light gluinos with mass $m_{\tilde g}= 3(5)$~GeV 
with 93(91)\% CL, while the mass range $\leq 1.5$~GeV remained
essentially unconstrained. Combining these results with the determination of
QCD colour coefficients from the analysis of multi-jet
events in~\cite{aleph}, the conclusions of \cite{Cs97} became much
stronger: light gluinos with mass $\leq 5$~GeV were excluded with at
least 99.89\% CL. The analysis of multi-jet events relied however
on the use of Monte Carlo simulations which parameters are tuned
to QCD without light gluinos. Moreover, the multi-jet analysis was based
on a tree-level calculation with rather large scale ambiguities.
The assessment of these uncertainties is difficult, thus preventing the
definite exclusion of a very light gluino by this argument~\cite{Fa97,pdg}.

Direct accelerator limits for the gluino as LSP were discussed
recently in Refs.~\cite{Ra99,Ba99,Ra00}: The authors of
Ref.~\cite{Ba99} concluded that the range 
$3~{\rm GeV}\lsim m_{\tilde g} \lsim$ 130--150~GeV 
can be excluded at 95\% C.L. based on currently available OPAL and CDF
data. Their results are sensitive to the details of the hadronic
interactions of $\tilde g$-hadrons and, for certain choices of the
parameters, a window in the intermediate mass
region $23~{\rm GeV}\lsim m_{\tilde g} \lsim 50~{\rm GeV}$ remains open.
Meanwhile, Ref.~\cite{Ra99} noted that these limits
could be weakened if squarks are not very heavy and contribute to the
jet + missing energy signal, while Ref.~\cite{Ra00} confirmed an open
window for a gluino with $25~{\rm GeV}\lsim m_{\tilde g} \lsim 35~{\rm GeV}$.

We also mention here that cosmological constraints do not exclude both
light and heavy gluinos of interest~\cite{Be98,Mo98}.

The Gustafson experiment \cite{gustaf} does not exclude 
$\tilde{g}$-hadrons (see section 5).

Till now we discussed the limits on the gluino mass $m_{\tilde g}$. 
The lightest $\tilde{g}$-hadron with mass $M_{\tilde g}$ is heavier
than the gluino by the mass of its constituent 
gluon or quarks, which are expected to be less than 1~GeV.

\begin{figure}
\vskip1cm
\begin{picture}(11,9)
 \put (3,0) {
 \epsfig{file=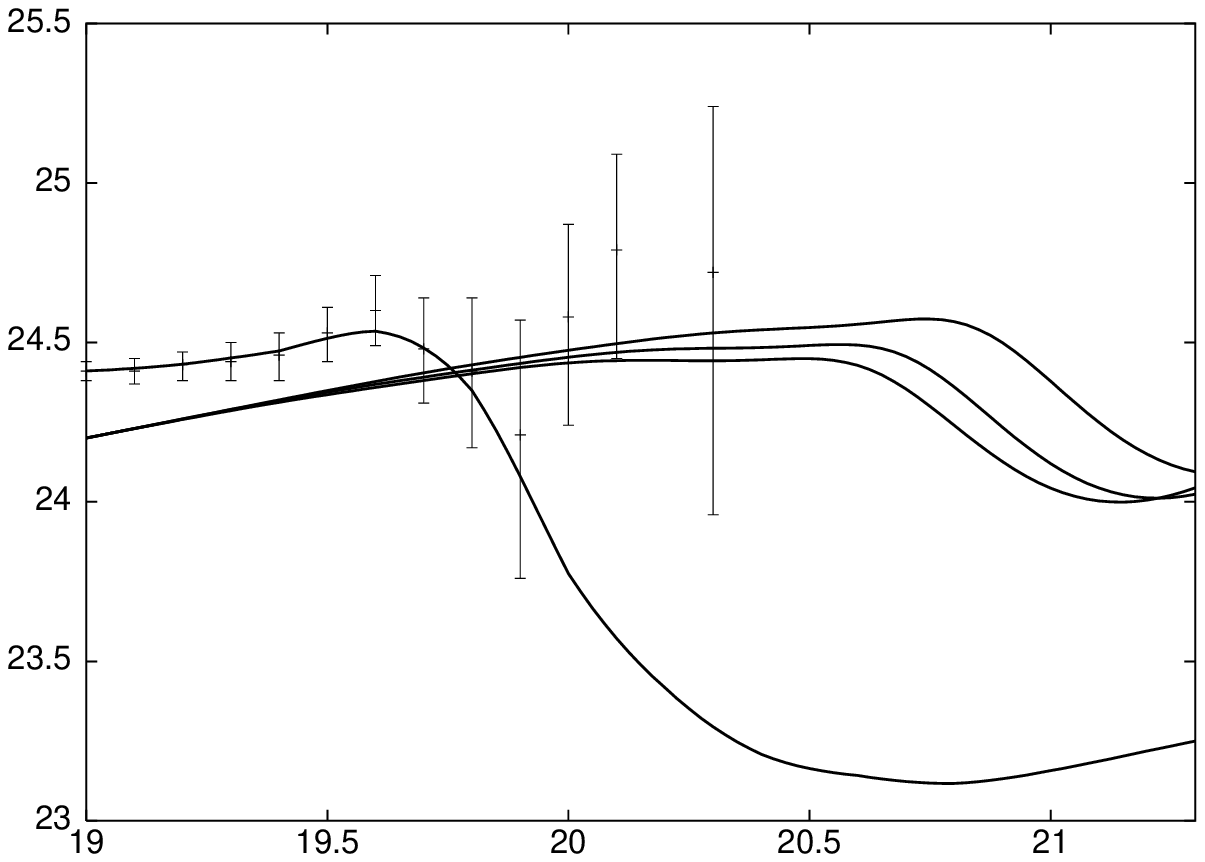,height=9.0cm,width=11.0cm}
 \put (-5.0,1.5) {proton}
 \put (-4.5,5.1) {1.5 GeV}
% \put (-3.2,7.0) {2 GeV}
 \put (-3.0,8.2) {3 GeV}
 \put (-9.1,-0.6) {$\log(E/{\rm eV})$}
 \put (-16.7,3.8) {\begin{sideways}
                   {$\log(E^3 J(E))/$m$^{-2}$s$^{-1}$eV$^{2}$}
                   \end{sideways}}    }
\end{picture}
\vskip0.4cm
\caption{\label{spec_Einf}
  Diffuse glueballino flux from uniformly distributed sources with
  injection spectra $\d N/\d E\propto E^{-2.7}$ for $M_{\tilde G}=1.5$, 
  2, and 3~GeV, compared with proton flux and observational data from
  AGASA. Either the gluino flux or a combination of proton and gluino
  fluxes can fit the data.}  
\end{figure}

Light $\tilde{g}$-hadrons  with $M_{\tilde g}\sim 1.5$~GeV
have a spectrum with the GZK cutoff beyond the currently observed
energy range (see \cite{Be98} and Fig.~\ref{spec_Einf} of the present
paper). Together with the accelerator limits on gluino
masses this leaves a narrow band of allowed masses for the light
$\tilde{g}$-hadrons at $1.5\lsim M_{\tilde g} \lsim 4~$GeV. But this
window is closed if, as argued  
in \cite{VO}, the charged gluebarino $\tilde{g}uud$ is lighter than 
the neutral $\tilde{g}uds$. Indeed, production of charged gluebarinos
in the Earth atmosphere by cosmic rays and their accumulation in the ocean    
results in too high abundance of ``wild hydrogen'' in contradiction
with  observational data. In Refs.~\cite{Fa96,Bu85},
however, it is argued that the lightest gluebarino is the neutral 
flavour singlet 
$\tilde{g}uds$, due to strong quark attraction in this state. 
But even in this case the restriction \cite{VO} might work, if 
$\tilde{g}uds$-gluebarino and proton are bound into anomalous deuterium.

In conclusion, a light gluino---although being disfavoured by
various arguments---is not excluded. We are studying here the interactions
of gluinos, being inspired by a possible correlation between AGN
and the arrival directions of UHECR~\cite{Fa98,corr} and by the recent  
suggestion \cite{Ch98,Al98},
that Extensive Air Showers (EAS) observed at the highest energies
could be produced by $\tilde g$-hadrons.
The authors of Ref.~\cite{Al98} performed a detailed simulation of EAS
induced by $\tilde g$-hadrons, using however a
phenomenological description for the $\tilde g$-hadron-nucleon
interaction which is not self-consistent. They found that masses as
high as 50~GeV are compatible with presently available data.
In contrast, it was argued in Ref.~\cite{Be98}, using kinematical
arguments, that the observed shower characteristics exclude any strongly
interacting particle much heavier than a few GeV.

The purpose of the present work is to study EAS produced by 
$\tilde g$-hadrons and  to restrict the mass range in
which the $\tilde g$-hadron is a viable UHE primary using a
self-consistent interaction model. We have used for the simulation of
air showers initiated by $\tilde g$-hadrons a suitably modified
version of the QGSJET  model~\cite{kalm97,izv94} 
which is known to describe successfully proton air
showers \cite{kalm97,fzk}. Specifically, we
have considered the glueballino as the $\tilde{g}$-hadron  but we
expect that our results apply to all $\tilde g$-hadrons. 
We paid special attention to consistent calculations of 
glueballino-hadron interaction cross-sections
and of cascade particle production in the atmosphere. 
We have found that the development of showers 
initiated by $\tilde g$-hadrons with masses above 5~GeV differs
substantially from proton-initiated showers and is inconsistent with
the current experimental data. In the window of masses 1.5--4~GeV, 
where a $\tilde{g}$-hadron can be allowed,
glueballino-induced showers do not contradict the available data. 
Future observations by the detectors HiRes and Auger can either
confirm or exclude $\tilde g$-hadrons as the dominant primary
combining the information from shower profiles and energy spectrum. 
However, the best way to test this hypothesis is a modified Gustafson 
experiment (see Section 5). In the case of the discovery of light
$\tilde{g}$-hadrons in such an experiment we shall reliably know their 
properties, thus enabling us to calculate the production of these
particles in astrophysical sources and their detection in the Earth
atmosphere. Their absence will preclude further discussion of this
hypothesis.

\section{Glueballino--proton (nucleus) interaction}
\subsection{QGSJET framework}
QGSJET, a Monte Carlo generator of hadron-hadron, hadron-nucleus, and
nucleus-nucleus interactions \cite{kalm97,izv94}, was developed in the
framework of the Gribov-Regge approach and is based on the quark-gluon
string model of the supercritical Pomeron \cite{kaid82}. Hadronic
interactions are described as a superposition of elementary rescattering
processes between the  partonic constituents of the
projectile and target nucleons (hadrons), resulting in the production
of colour neutral strings, which further fragment into secondary
hadrons. The key parameters of the model are the intercepts and
the slopes of the Regge trajectories of the Pomeron and of secondary
Reggeons. These parameters govern the formation of different interaction
configurations, how the energy-momentum is shared in elementary
interactions, and also the string hadronization.
The model was generalized to hadron-nucleus and nucleus-nucleus
interactions in the framework of the Glauber-Gribov
approach~\cite{glaub,grib}, taking into account low mass diffraction
and inelastic screening processes \cite{yad93}. 
Hard QCD processes were included into the Gribov-Regge formalism 
via the concept of a "semihard Pomeron", which is a $t$-channel
iteration of the soft Pomeron and a QCD parton ladder
contribution \cite{izv94,proc97}.

QGSJET describes hadron-hadron interaction amplitudes as a sum
of two contributions, namely soft and semihard rescatterings \cite{izv94}. 
The soft contributions are of purely nonperturbative nature and
correspond to the case of a parton cascade with virtualities smaller
than some cutoff $Q_0^2$. Below this cutoff, perturbative QCD is not  
applicable and the interaction is described by phenomenological soft
Pomeron exchange. 
The amplitude $f_{ac}^{\P}$ for Pomeron exchange between two hadrons $a$
and $c$ is given by \cite{kaid82}
\begin{eqnarray}   \label{softpom}
 f^{\P}_{ac}\! \left(s,b\right)  & = & 
 \frac{\gamma_{a}\gamma_{c}\,\exp \!(\Delta y)}{\lambda_{ac}(y)}
 \exp \! \left( -\frac{b^{2}}{4\lambda _{ac}(y)}\right)
 \\  \label{lambda}
 \lambda_{ac}(y) & = & R_{a}^{2}+R_{c}^{2}+ \alpha_{\P}'y \,,
\end{eqnarray}
where $y=\ln s$ is the rapidity size of the Pomeron, $s$ is the
squared center-of-mass energy  for the interaction,
 $b$ is the impact parameter between the
two hadrons,  and the parameters $\gamma_{a(c)}$, $R^2_{a(c)}$ are the
vertices and slopes for the Pomeron-hadron $a(c)$ coupling, respectively.
Finally, $\Delta$ and $\alpha_{\P}'$ are the parameters describing  
the overcriticality and the slope of the soft Pomeron trajectory.

Contrary to soft rescatterings, semi-hard ones correspond to the case
when at least a part of the parton cascade develops in the region of
parton virtualities $q^2>Q_0^2$ and, therefore, can be described on the 
basis of QCD techniques. The complete semi-hard contribution is represented
by a QCD parton ladder sandwiched between two soft Pomerons \cite{proc97}. 
For the Pomeron, the formulas (\ref{softpom}-\ref{lambda}) can still be used.
However, since it is now coupled to a hadron $a(c)$ on one
side but to a parton ladder on the other side,  
the slope $R^2_{c(a)}$ and the coupling $\gamma_{c(a)}$ have to be
replaced by the slope $R^2_{\rm lad}$ and the coupling  $V^{\P}_{\rm lad}$
of the Pomeron-ladder. The latter is parameterised as
\begin{equation}
 V^{\P}_{\rm lad}(y)=r\left[1-\exp \!(-y)\right]^{\beta_{\P}},
\end{equation}
where the parameters $r$ and $\beta_{\P}$ describe the momentum
distribution of a parton (sea quark or gluon) in the soft Pomeron. Using  
$R^2_{\rm lad}\simeq 1/Q_0^2 \ll R^2_{a(b)}+\alpha_{\P}'y$, the 
slope $R^2_{\rm lad}$ can be neglected.

To complete the description of soft and semihard contributions, the
momentum distribution function $N^{\P}_a$ for soft Pomeron emission by a
hadron of type $a$ has to be specified. It is parameterised in the form 
 \begin{equation}  \label{N}
 N^{\P}_{a}\!\left(x_{\P}^{\pm}\right) \sim 
  \left(x_{\P}^{\pm}\right)^{\alpha}
  \left(1-x_{\P}^{\pm}\right)^{\beta_{a}} \,, 
\end{equation}
where the first factor does not depend on the hadron type and describes
the probability to slow down the hadron constituents to which the
Pomeron is connected. In QGSJET, the Pomeron is connected to a
(dressed) quark-antiquark pair. Using Regge asymptotics \cite{kaid82}, i.e. 
$\alpha_{q\bar q}\simeq 0.5$ as the intercept of the Regge $q\bar
q$-trajectory for light quarks, it follows  
\begin{equation}
 \alpha=1-2\alpha _{q\bar q}\simeq 0 \,.
\end{equation}
Similar, the second factor in Eq.~(\ref{N}) describes the probability to 
slow down the "leading" hadron state configuration; the parameter
$\beta_{a}$ is expressed via the intercept $\alpha_{a\bar a}$ of the
corresponding Regge trajectory as $\beta_{a}=-\alpha_{a\bar a}$.

The semihard contribution described above corresponds to the case that 
gluons or sea quarks of the hadron start the interaction at the
initial scale $Q_0^2$. Additionally, valence quarks can interact with
$q^2\geq Q_0^2$. Then the only nonperturbative input needed are the
valence quark momentum distributions $q_v\left(x,Q_0^2\right)$ at the
initial scale $Q_0^2$.

\subsection{Extension for glueballino}
\label{qgsjet}
The nucleon-glueballino interaction can be treated in the QGSJET model in
the same framework as the one for usual hadrons \cite{kalm97}. The main
difficulty is to connect the unknown physical parameters 
(coupling $\gamma _{\tilde G}$, slope $R^2_{\tilde G}$ and momentum
distribution  $N_{\tilde G}^{\P}$) describing the
interactions of glueballinos with the corresponding 
quantities of usual hadrons.
We use simple scaling arguments to derive the glueballino
parameters from those of the pion:

\begin{enumerate}
\item  
The coupling $\gamma_a$ of a hadron $a$ to the Pomeron depends
essentially on its size and, consequently, on its reduced mass
$M_a$. If $r_a$ denotes the radius of the hadron $a$ with the reduced
 mass $M_a$, then
$\gamma_a\sim r^2_a\sim M_a^{-2}$, where we have neglected a factor
$\alpha_s(M_a^2)$ in the last step. Thus, the Pomeron-glueballino vertex
$\gamma _{\tilde G}$ can be expressed via the Pomeron-pion vertex
$\gamma _{\pi}$ as 
\begin{equation}  \label{gammaglu}
 \gamma _{\tilde G}=\gamma _{\pi}\left(\frac{M_{\pi}}
 {M_{\tilde G}}\right)^2 \,.  
\end{equation}
For the reduced mass $M_{\tilde G}$ of the glueballino, we use
\begin{equation}
 M_{\tilde G}=\frac{m_{\tilde g}m_g}{m_{\tilde g}+m_g} \,,
\end{equation}
where $m_{\tilde g}$ is the mass of the gluino and $m_g\simeq 0.7$~GeV is
the constituent mass of the gluon. Similar, we use for the pion
$M_{\pi}= m_q/2$ with $m_q\simeq 0.35$~GeV as quark constituent mass.
Note that Eq.~(\ref{gammaglu}) does not take into account the 
different colour factors of quarks and gluon/gluinos because we consider
an effective Pomeron coupling to the hadron as a whole, not to 
individual parton constituents.
\item
The slope $R^2_{\tilde G}$ for the Pomeron-glueballino coupling is also
inverse proportional to $M_{\tilde G}^2$. Therefore, $R^2_{\tilde G}$ 
is small compared to $R^2_{p}$ and $R^2_{\P}=\alpha_{\P}'y$
and can be neglected in the formulas (\ref{softpom}-\ref{lambda}),
\begin{equation}
 \lambda _{\tilde Gp}\!(y) \simeq R^2_{p}+\alpha_{\P }'y \,.
\end{equation}
\item
The momentum distribution for Pomeron emission is again given by
Eq.~(\ref{N}), \mbox{$N^{\P}_{\tilde G}\sim$} 
              \mbox{$(x_{\P}^\pm)^\alpha (1-x_{\P}^\pm)^{\beta_{\tilde G}}$.}
Now the "leading" configuration consists of a valence gluon and
gluino, 
\begin{equation}
 \beta _{\tilde G}=1+\beta _{\tilde g}+\beta _{g} \,.
\end{equation}
Assuming that a valence gluon behaves similar to a valence $q\bar q$-pair in
the low $x$-limit gives $\beta_{g}\simeq 1-2\alpha _{q\bar q}\simeq 0$. The
remaining unknown parameter $\beta_{\tilde g}$ can be found from the
momentum distribution between the valence constituents of the 
glueballino. Using as ansatz for the momentum distribution
$\rho_{\tilde g }^{\tilde G}$ of the gluino 
\begin{equation}  \label{glumom}
 \rho_{\tilde g }^{\tilde G}\left(x_{\tilde g }\right) \sim 
 x_{\tilde g }^{\beta _{\tilde g}}
 \left(1-x_{\tilde g }\right)^{\beta_{g}} 
\end{equation}
and assuming that the energy is shared according to the constituent
masses of the valence partons, we obtain for the average
momentum fraction carried by the gluino
\begin{equation}  \label{xglu}
\langle x_{\tilde g }\rangle =\frac{m_{\tilde g}}{m_{g}+m_{\tilde g}}=
\frac{\beta _{\tilde g}+1}{\beta _g+\beta _{\tilde g}+2}  \,.
\end{equation}
This results in
\begin{equation}
 \beta _{\tilde g} =\frac{m_{\tilde g}}{m_{g}}\left(\beta _g+1\right)-1
 \simeq 
 \frac{m_{\tilde g}}{m_{g}}-1 \,.
\end{equation}
\end{enumerate}

Having fixed the free parameters describing the Pomeron-glueballino
interactions using essentially only one simple, physically
well-motivated scaling argument, the soft and semi-hard contributions 
are determined. 
These two contributions to the total nucleon-glueballino interaction
are referred below as the contribution due to the
{\em soft coupling\/}, because they are both caused by soft Pomeron emission
of the glueballino. 

To complete the formalism, we need to define the momentum distributions
of the valence gluon or gluino inside the glueballino probed at the
initial scale $Q^2_0$, when they are involved into hard
interactions. We shall refer to this contribution below as the
the contribution due to the {\em direct coupling}. Parton emission by a
gluino in the $s$-channel is strongly suppressed kinematically 
in the nonperturbative region $q^2<Q_0^2$ by its mass: the virtuality
$q^2$ of the  process $\tilde g \to g+\tilde g$ is determined by the 
off-shellness $\left|q^2_{\tilde g}-M^2_{\tilde g}\right|$
of the produced $t$-channel gluino, 
\begin{equation}   \label{kinem}
 q^2_{\tilde g\rightarrow \tilde g} =
 \left|q^2_{\tilde g}-M^2_{\tilde g}\right| =
 \frac{p_{\perp}^2}{1-z}+ M^2_{\tilde g} (1-z)  \,.
\end{equation}
Here $p_{\perp}$ and $z$ are the  transverse momentum and the light 
cone momentum fraction for the $t$-channel gluino. 
Therefore, the gluino momentum distribution at the scale $Q_0^2$
essentially remains in the form (\ref{glumom}),
\begin{equation} \label{gluv}
 \tilde g_v\!\left(x_{\tilde g },Q_0^2\right)=
 \rho_{\tilde g }^{\tilde G}\left(x_{\tilde g }\right) \,.
\end{equation}
By contrast, the momentum distribution of the valence gluon differs
from $\rho_{\tilde g }^{\tilde G}\left(1-x_g\right)$ in (\ref{glumom})
because of the emission of sea quarks and gluons (due to the soft
coupling defined above), 
\begin{equation}
 g_v\!\left(x_{ g },Q_0^2\right)\sim
 x_{ g }^{\beta _{ g}}
 \left(1-x_{g }\right)^{\beta _{\tilde g }+\delta} \,.
\end{equation}
The parameter $\delta$ is fixed by the momentum conservation
constraint for the complete (valence and sea) parton momentum
distributions at the initial scale $Q^2_0$. This
completes the formulation of the initial conditions for the
perturbative evolution. 

The treatment of the perturbative part of the interaction is performed
to leading-log accuracy, the corresponding techniques are described
in Ref.~\cite{kalm97}. 
All emitted ($s$-channel) partons undergo timelike cascading
according to the standard algorithm \cite{webber}, with soft gluon
coherence taken into account. At the final stage, soft strings are assumed
to be formed between on-shell partons according to their
colour connection pattern. The final gluino is assumed to pick up a
gluon-gluon singlet pair from the vacuum. After a colour rearrangement
similar to the one in $J/\psi$ production, a glueballino is formed.
String hadronization completes the procedure for glueballino-hadron
interaction.
The extension of the model to glueballino-nucleus collisions is based
on the standard Glauber-Gribov approach and does not differ
from the usual hadron-nucleus case \cite{yad93}.

\subsection{Numerical results}
The model developed in the last subsection allows both to calculate
the cross-sections for glueballino-nucleon interactions and to treat
consistently particle production in these reactions. Some quantities
characterising the glueballino-nucleon interactions are
given in Table~1 for $E_{\rm lab}=100$~GeV and in Table~2 for
$E_{\rm lab}=10^{12}$~GeV. We present both total and inelastic
cross-sections as well as the partial contributions arising due to the
soft ($\sigma_{\rm tot}^{\rm s-coupl.}$) and direct 
($\sigma_{\rm tot}^{\rm d-coupl.}$) coupling of the glueballino%
\footnote{Note that the total cross section is smaller than the sum of
  the partial contributions $\sigma _{\rm tot}^{\rm s-coupl.}$ and 
  $\sigma_{\rm tot}^{\rm d-coupl.}$, because it contains contributions from
  multiple rescatterings of both types.}.
At low energies, the soft coupling strongly dominates  the 
$\tilde G$-proton interaction for all $M_{\tilde G}$
being governed by  nonperturbative soft interactions, 
while the direct coupling can be neglected. 
At high energies, this picture changes considerably.
The soft coupling becomes more and more suppressed for large $M_{\tilde G}$.
By contrast, the direct contribution, which is purely perturbative on
the glueballino side, is nearly independent of $M_{\tilde G}$. 
This important difference from the usual
hadron case is due to the very asymmetric energy partition between parton
constituents of the glueballino. For large $M_{\tilde G}$, 
the valence gluino carries almost
the whole initial energy of the particle (88\% for $M_{\tilde g}=5$ GeV
and 99\% for $M_{\tilde g}=50$ GeV) -- Eqs. (\ref{glumom}-\ref{xglu}),
 (\ref{gluv}), thus leaving just a small part of it to other partons, 
 to which Pomerons are connected.
Therefore, the glueballino behaves in the limit of large $M_{\tilde G}$ 
essentially as a perturbative object, as one expects from
kinematical considerations \cite{Be98}.
Finally, the last raw of the tables shows the inelasticity coefficient
$K_{\rm inel}$ as another important quantity which distinguishes
proton-proton and $\tilde G$-proton interactions. Although at energies
of interest for UHECR the total cross-section for $\tilde G$-proton
interactions is rather large, a heavy glueballino behaves like a
penetrating particle in the atmosphere loosing only a small part of
its energy in one interaction. This conclusion was already reached
in Ref.~\cite{Be98} from semiqualitative considerations.
The reason for this effect is twofold. On one side, as discussed
above, gluinos of larger masses carry a
larger fraction of the initial particle energy, leaving a smaller part of it
for the sea constituents ((anti-)quarks and gluons) and thus reducing the
average number of multiple interactions in $\tilde G$-proton (nucleus)
collisions. 
On the other hand, the relative weight of the "direct hard" process
increases for heavier gluinos,  where the valence gluino
loses typically only a small part of its energy, as a large longitudinal
momentum transfer is strongly suppressed in that case by the process
virtuality $q^2_{\tilde g\rightarrow g}$,
\begin{equation}   \label{kinem-t}
 q^2_{\tilde g\rightarrow g}=\frac{p_{\perp}^2}{1-z}+
 \frac{z^2M_{\tilde g}^2}{1-z}  \,,
\end{equation}
with  $p_{\perp}$ and $z$ being the  transverse momentum and the light 
cone momentum fraction for the  $t$-channel gluon emitted of the initial
valence gluino which mediates the gluino hard interaction with the target
proton (nucleus). 

\begin{figure}
\begin{picture}(11,9)
 \put (3,0) {
 \epsfig{file=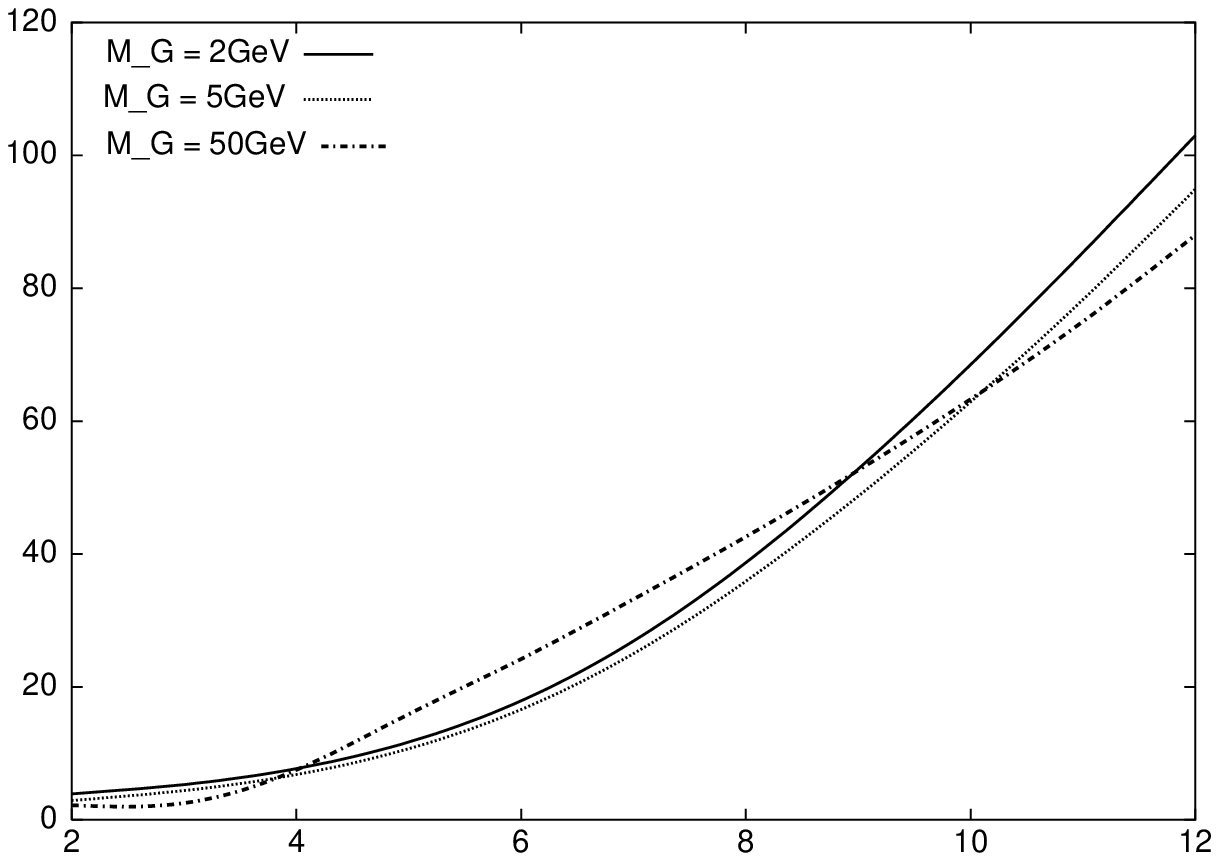,height=9.0cm,width=11.0cm}  
 \put (-9.1,-0.6) {$\log(E_{\rm lab}/{\rm GeV})$}
 \put (-16.7,7.0) {{\Large $\sigma_{\rm tot}$}}      }
\end{picture}
\vskip0.6cm
\caption{\label{tot}
         Total glueballino-nucleon cross-section $\sigma_{\rm tot}$ in
         mbarn as function of $E_{\rm lab}$ for $M_{\tilde G}=2,5$ and
         50~GeV.}
\end{figure}

\begin{figure}
\vskip-2.0cm
\begin{picture}(11,14)
 \put (3,0) {
 \epsfig{file=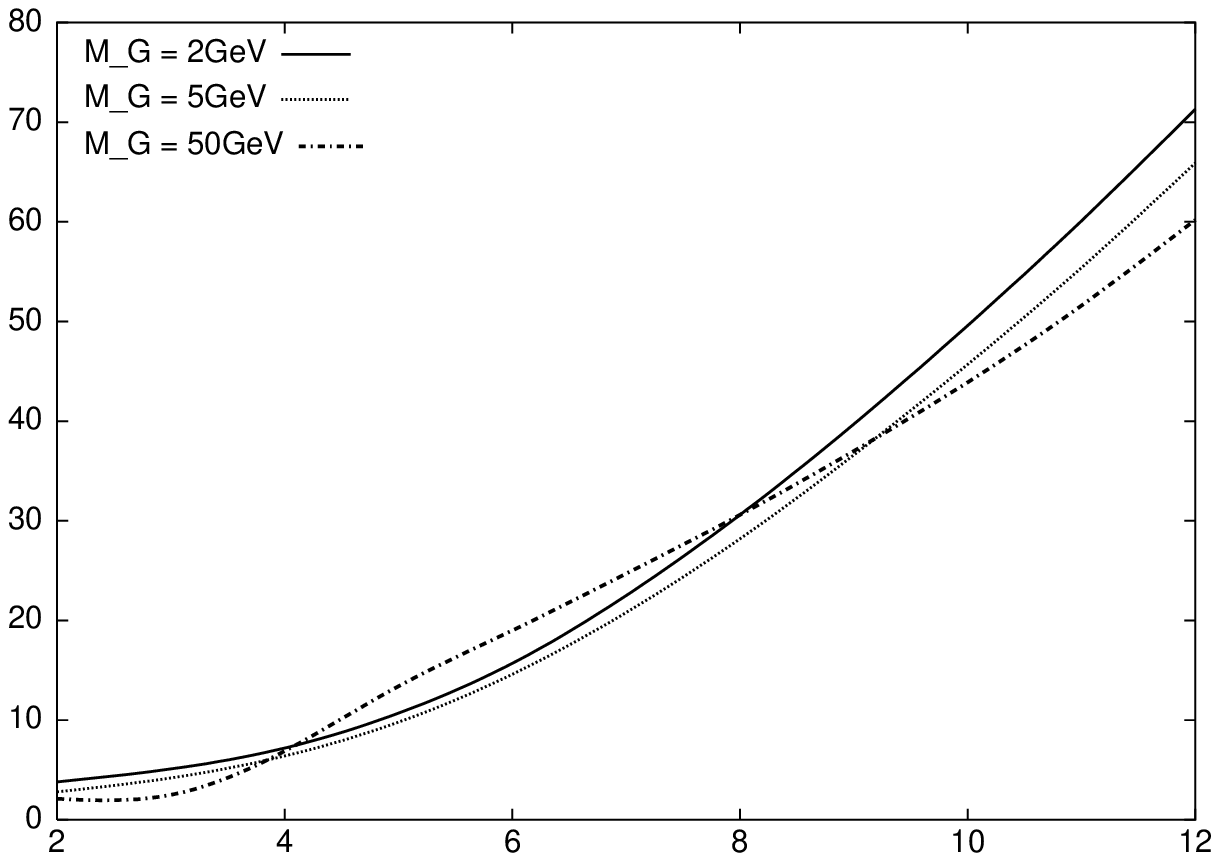,height=9.0cm,width=11.0cm}   
 \put (-9.1,-0.6) {$\log(E_{\rm lab}/{\rm GeV})$}
 \put (-16.7,7.0) {{\Large $\sigma_{\rm in}$}}      }
\end{picture}
\vskip0.6cm
\caption{\label{inel}
         Inelastic glueballino-nucleon cross-section $\sigma_{\rm in}$
         in mbarn as function of $E_{\rm lab}$ for $M_{\tilde G}=2,5$
         and 50~GeV.} 
\end{figure}

We show also in Figs.~\ref{tot} and \ref{inel} the total
and inelastic cross-sections as function of the interaction energy
$E_{\rm lab}$ for
glueballino masses $M_{\tilde G}=2,5$ and $50$~GeV.
The fast increase of the direct contribution with 
 $E_{\rm lab}$ produces an interesting effect: the
total interaction cross-section for the largest glueballino mass 
considered, $M_{\tilde G}=50$~GeV, which is dominated by the direct
contribution, overshoots the ones for smaller 
glueballino masses in the energy range $10^4-10^9$~GeV.

\section{Extensive air showers (EAS)}
\begin{figure}
\begin{picture}(11,9)
 \put (3,0) {
 \epsfig{file=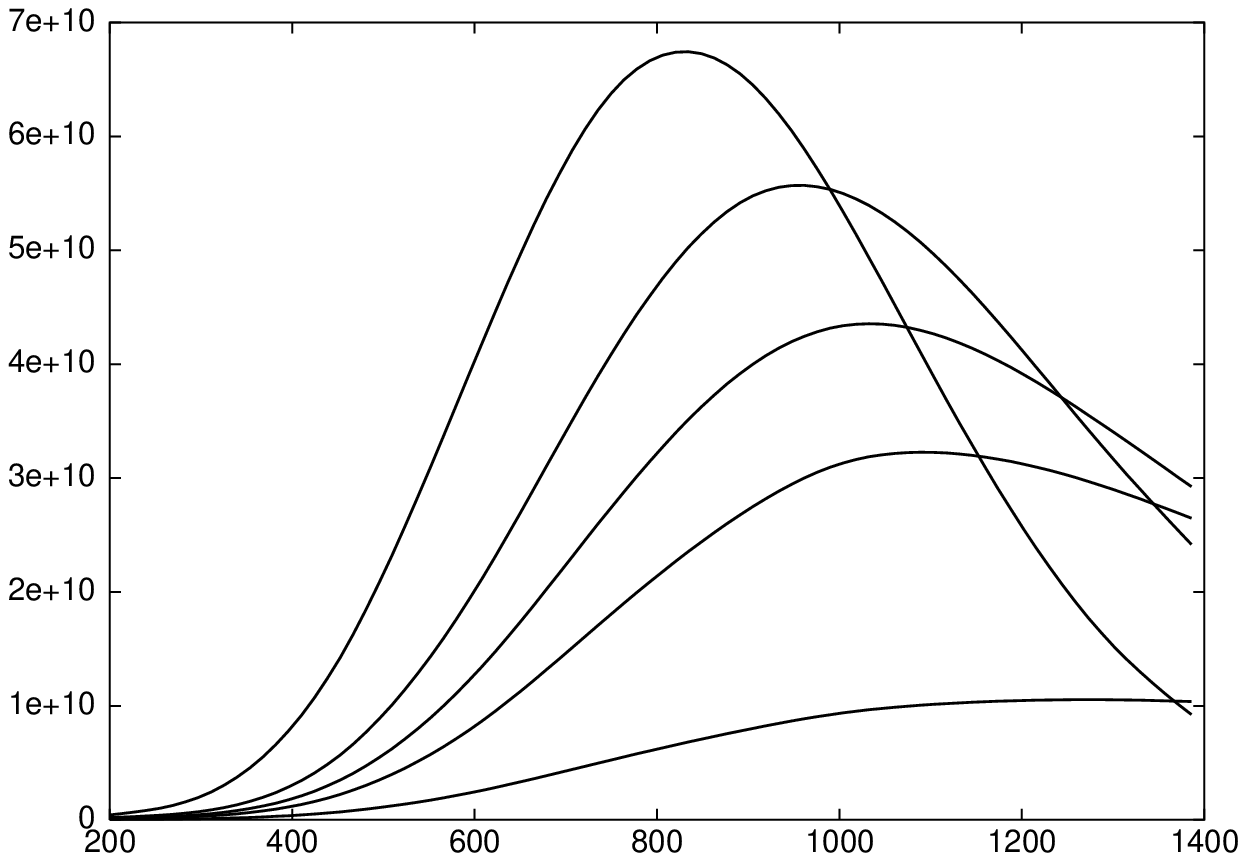,height=9.0cm,width=11.0cm} 
 \put (-11.0,10.0) {proton}
 \put (-4.2,9.4) {$\tilde G$(2 GeV)}
 \put (-7.7,6.8)   {$\tilde G$(5 GeV)}
 \put (-6.4,4.7)   {$\tilde G$(10 GeV)}
 \put (-6.0,2.7)   {$\tilde G$(50 GeV)}
 \put (-9.5,-0.6) {\large $X_{\max}$, g/cm$^2$}
 \put (-17.5,6.6) {\large $N_e$}                  }
\end{picture}
\vskip0.6cm
\caption{\label{pro_E20}
         Longitudinal shower profile for EAS of energy $E_0=10^{20}$~eV
         initiated by protons and by glueballinos with 
         $M_{\tilde g}=2,5,10$ and 50~GeV.}
\end{figure}

\begin{figure}
\vskip-2.0cm
\begin{picture}(11,14)
 \put (3,0) {
 \epsfig{file=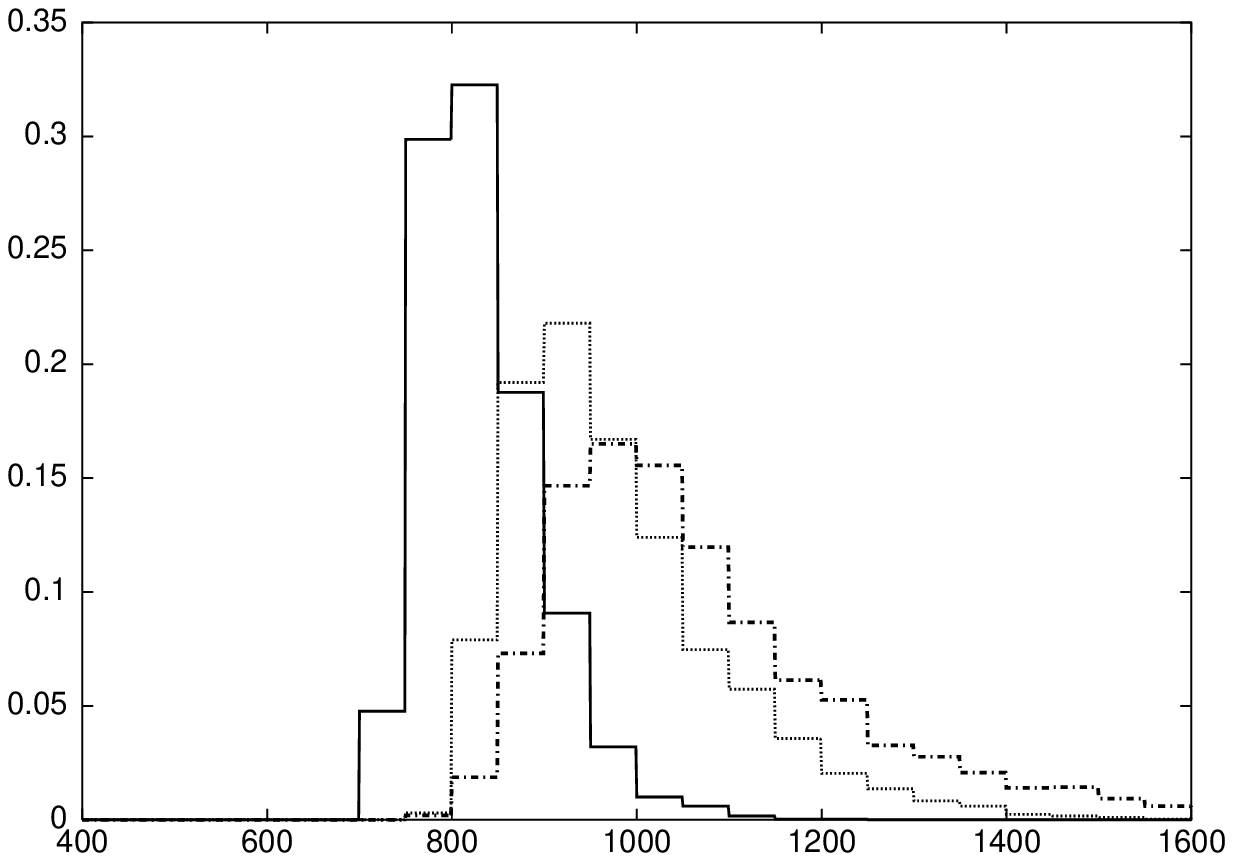,height=9.1cm,width=11.2cm}
 \put (-12.8,11.0) {proton}
 \put (-7.9,7.6) {$\tilde G$(2 GeV)}
 \put (-4.5,2.9)   {$\tilde G$(5 GeV)}
 \put (-9.5,-0.6) {{\large $X_{\max}$, g/cm$^2$}}
 \put (-18.5,6.6) {{\large $p(X_{\max})$}}         }
\end{picture}
\vskip0.6cm
\caption{\label{xmax_E20}
         Normalized distribution $p(X_{\max})$ of the shower maxima 
         for EAS  of energy $E_0=10^{20}$~eV
         initiated by protons and by glueballinos with 
         $M_{\tilde g}=2$ and 5~GeV.}
\end{figure}

\begin{figure}
\begin{picture}(11,9)
 \put (3,0) {
 \epsfig{file=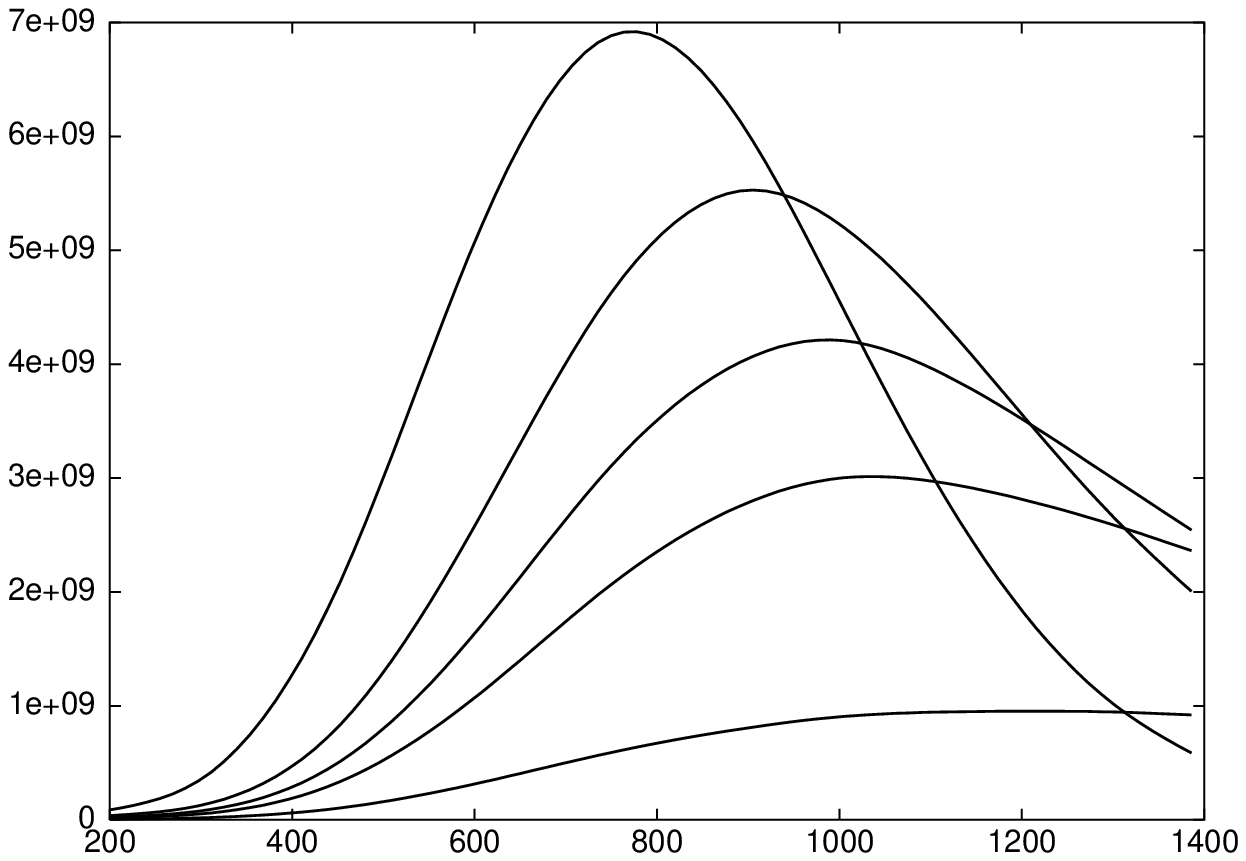,height=9.0cm,width=11.0cm} 
 \put (-11.5,10.0) {proton}
 \put (-4.2,9.0) {$\tilde G$(2 GeV)}
 \put (-7.9,6.8)   {$\tilde G$(5 GeV)}
 \put (-7.5,3.9)   {$\tilde G$(10 GeV)}
 \put (-6.0,2.5)   {$\tilde G$(50 GeV)}
 \put (-9.5,-0.6) {\large $X_{\max}$, g/cm$^2$}
 \put (-17.5,6.6) {\large $N_e$}                  }
\end{picture}
\vskip0.6cm
\caption{\label{pro_E19}
         Longitudinal shower profile for EAS of energy $E_0=10^{19}$~eV
         initiated by protons and by glueballinos with 
         $M_{\tilde g}=2,5,10$ and 50~GeV.}
\end{figure}

\begin{figure}
\vskip-2.0cm
\begin{picture}(11,14)
 \put (3,0) {
 \epsfig{file=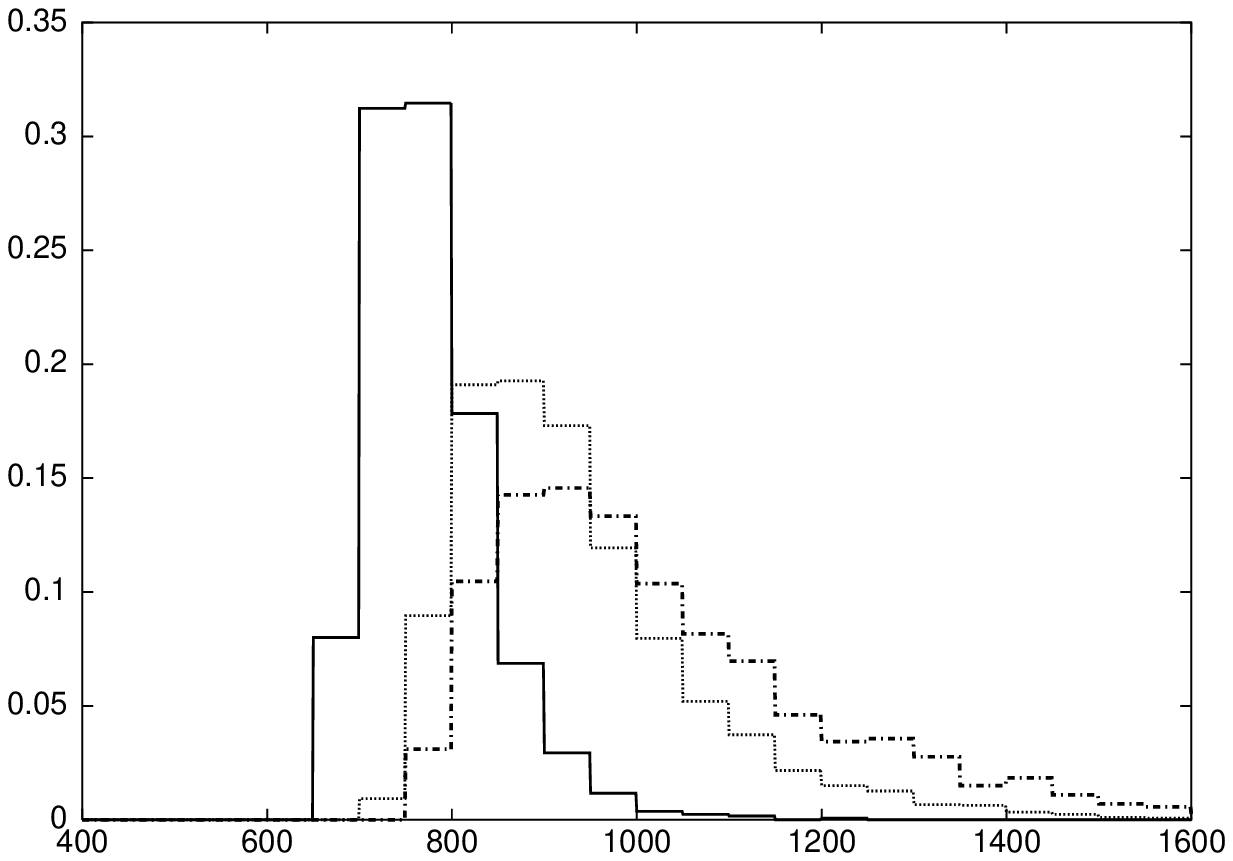,height=9.0cm,width=11.0cm}
 \put (-13.0,11.0) {proton}
 \put (-8.1,7.0) {$\tilde G$(2 GeV)}
 \put (-4.3,2.4)   {$\tilde G$(5 GeV)}
 \put (-9.5,-0.6) {{\large $X_{\max}$, g/cm$^2$}}
 \put (-18.5,6.6) {{\large $p(X_{\max})$}}         }
\end{picture}
\vskip0.6cm
\caption{\label{xmax_E19}
         Normalized distribution $p(X_{\max})$ of the shower maxima 
         for EAS  of energy $E_0=10^{19}$~eV
         initiated by protons and by glueballinos with 
         $M_{\tilde g}=2$ and 5~GeV.}
\end{figure}

\begin{figure}
\begin{picture}(11,9)
 \put (3,0) {
 \epsfig{file=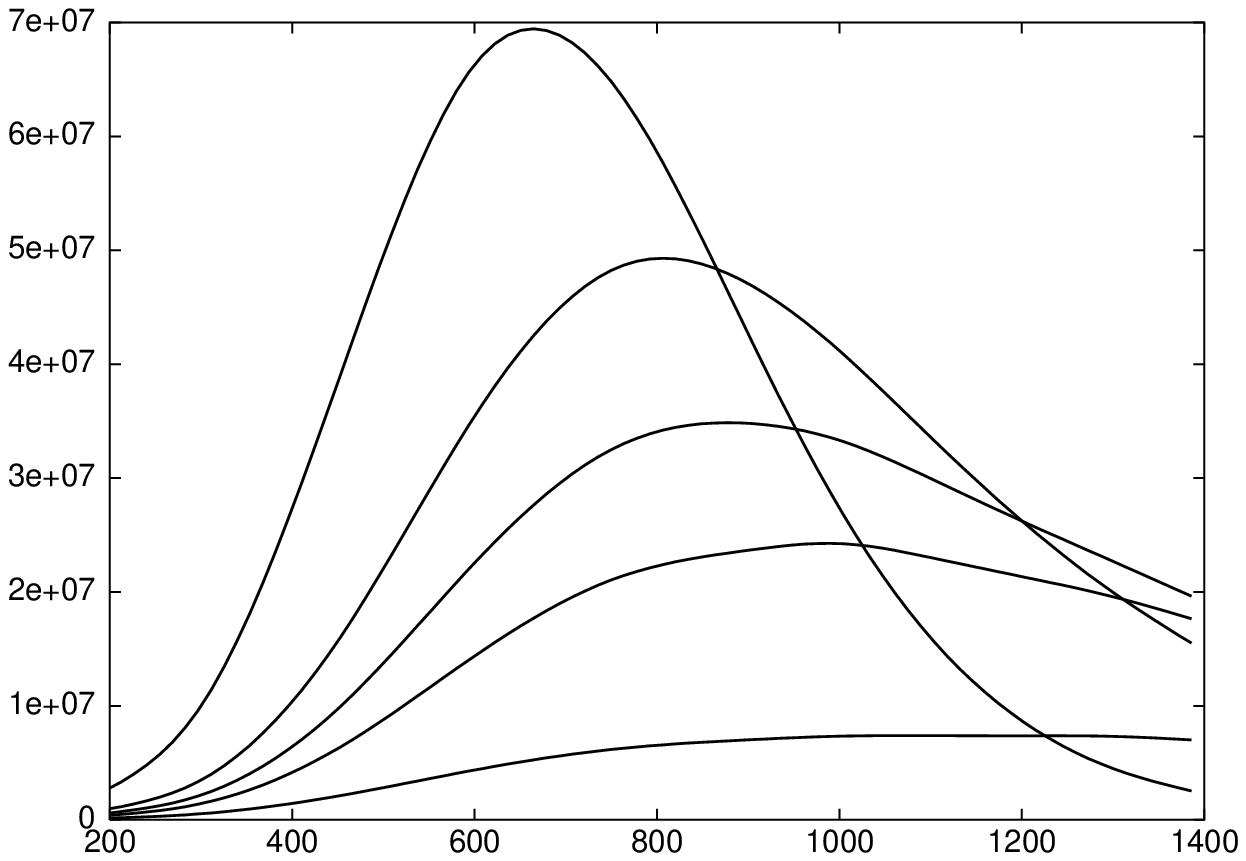,height=9.0cm,width=11.0cm} 
 \put (-12.0,11.0) {proton}
 \put (-5.2,8.0) {$\tilde G$(2 GeV)}
 \put (-8.7,6.8)   {$\tilde G$(5 GeV)}
 \put (-8.2,4.9)   {$\tilde G$(10 GeV)}
 \put (-7.5,2.2)   {$\tilde G$(50 GeV)}
 \put (-9.5,-0.6) {\large $X_{\max}$, g/cm$^2$}
 \put (-17.5,6.6) {\large $N_e$}                  }
\end{picture}
\vskip0.6cm
\caption{\label{pro_E17}
         Longitudinal shower profile for EAS of energy $E_0=10^{17}$~eV
         initiated by protons and by glueballinos with 
         $M_{\tilde g}=2,5,10$ and 50~GeV.}
\end{figure}

\begin{figure}
\vskip-2.0cm
\begin{picture}(11,14)
 \put (3,0) {
 \epsfig{file=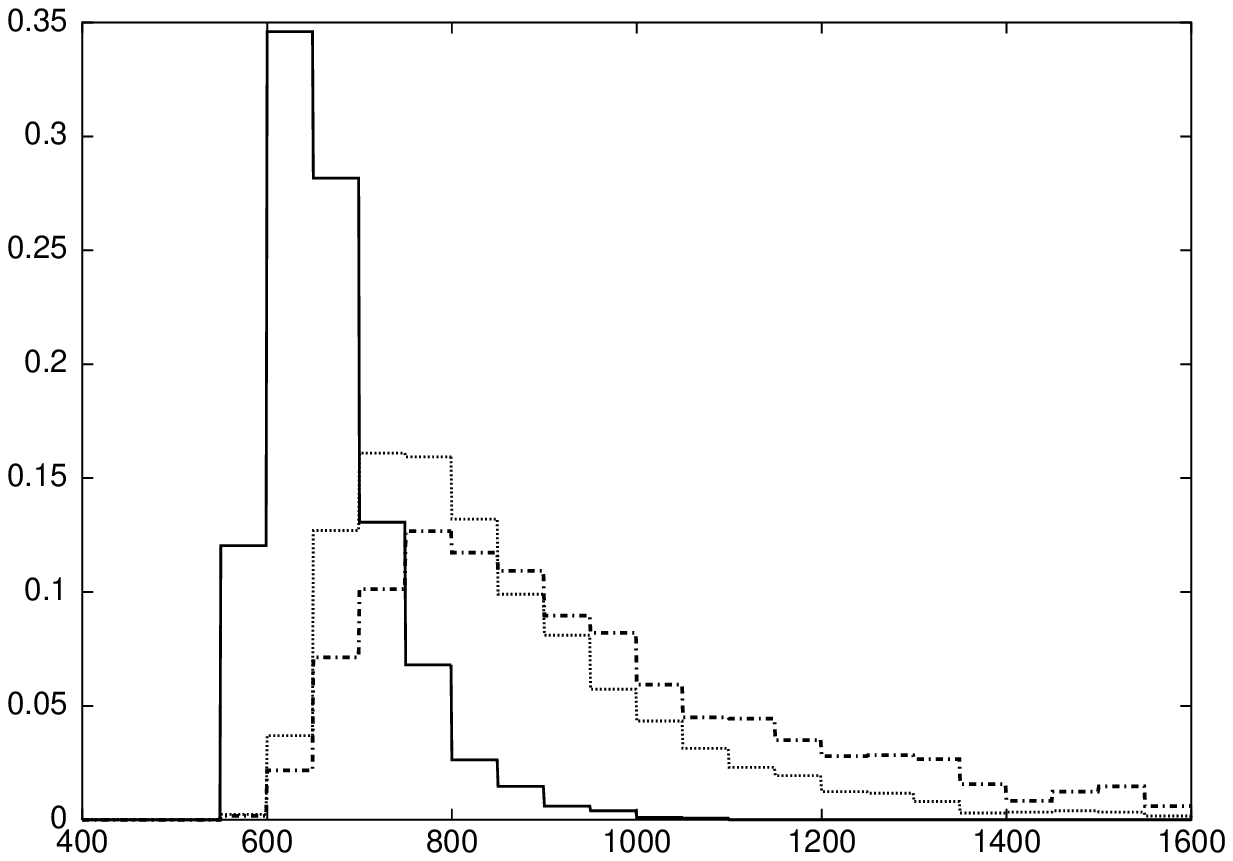,height=9.0cm,width=11.0cm}
 \put (-12.5,11.0) {proton}
 \put (-7.9,7.6) {$\tilde G$(2 GeV)}
 \put (-4.5,2.6)   {$\tilde G$(5 GeV)}
 \put (-9.5,-0.6) {{\large $X_{\max}$, g/cm$^2$}}
 \put (-18.5,6.6) {{\large $p(X_{\max})$}}         }
\end{picture}
\vskip0.6cm
\caption{\label{xmax_E17}
         Normalized distribution $p(X_{\max})$ of the shower maxima 
         for EAS  of energy $E_0=10^{17}$~eV
         initiated by protons and by glueballinos with 
         $M_{\tilde g}=2$ and 5~GeV.}
\end{figure}

\begin{figure}
\vskip2cm
\begin{picture}(11,9)
 \put (3,0) {
 \epsfig{file=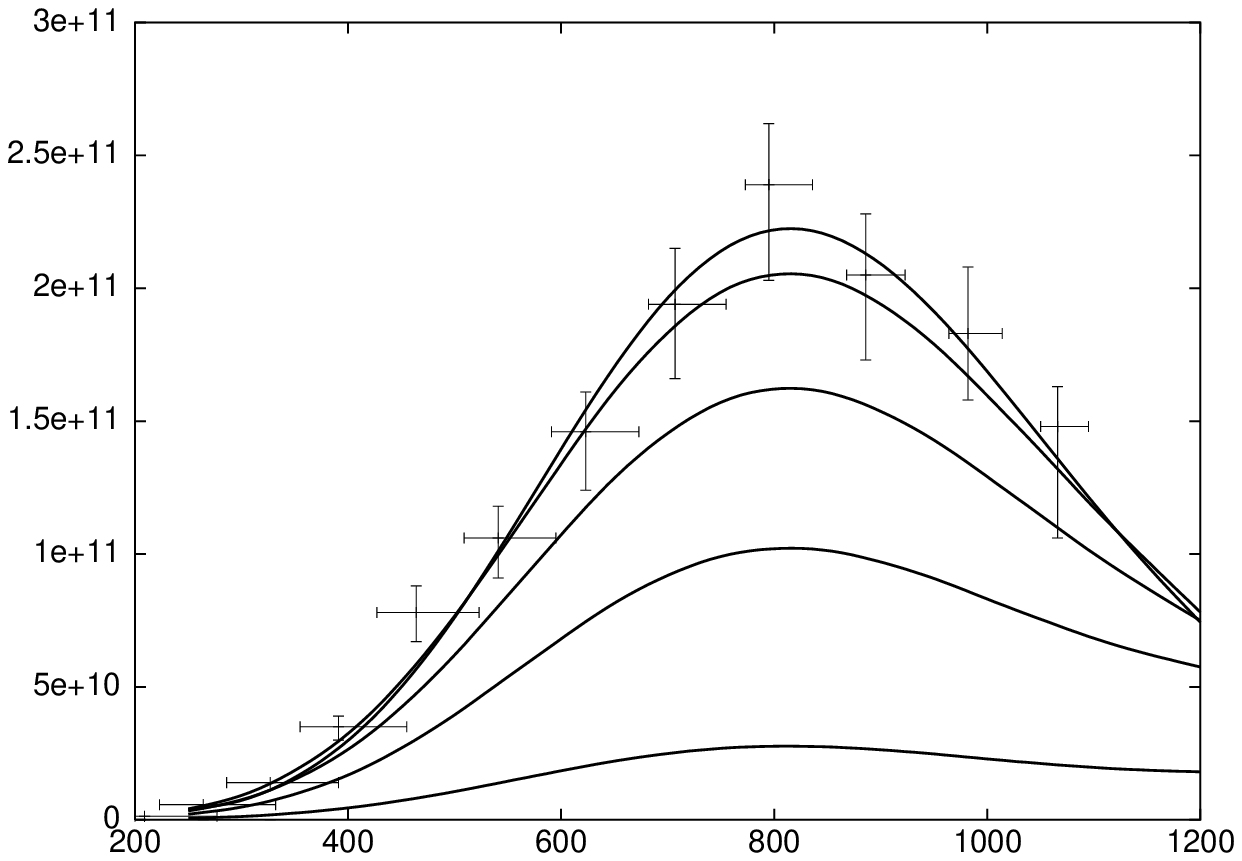,height=9.0cm,width=11.cm}
 \put (-8.0,-0.6) {{\Large $X_{\max}$, g/cm$^2$}}
 \put (-17.5,7.0) {{\Large $N_e$}}                    }
\end{picture}
\vskip0.6cm
\caption{\label{FE}
  Comparison of Flye's Eye highest energy event with the 
  longitudinal shower profile for EAS of energy $E_0=3\times 10^{20}$~eV
  initiated (from top to down) by protons and by glueballinos with 
  $M_{\tilde g}=2,5,10$ and 50~GeV. The shower profiles are shifted so that
  their $X_{\max}$ agrees with the observed shower maximum.}
\vskip12.0cm
\end{figure}

In this section, we present some results of our simulations for the
glueballino-initiated EAS. Shower profiles and distributions
of shower maxima are shown in Figs.~\ref{pro_E20}--\ref{xmax_E17} for three
different initial energies, $E_0=10^{17}$, $10^{19}$ and  $10^{20}$~eV,
for the glueballino as primary with different choices of the gluino
mass. For comparison, the case of a primary proton is also shown.
At the highest energy considered, $E_0=10^{20}$~eV,
the longitudinal shower profiles (Fig.~\ref{pro_E20}) and the distribution
of the shower maxima $X_{\rm max}$ (Fig.~\ref{xmax_E20}) of
glueballino-induced EAS are comparable with those induced by protons
in the case of glueballino masses smaller than 5~GeV.
As the glueballino mass increases, the shower develops deeper in the
atmosphere with a less pronounced maximum. The fluctuations in
$X_{\max}$ increase also for larger glueballino masses.
The  main reason for both effects is the
competition between the large glueballino-nucleus cross-section and the
small inelasticity of the interactions: a heavy glueballino injects
in one interaction only a small part of its energy into secondary
hadronic and electromagnetic cascades, while interactions with a large
momentum transfer are rare and increase only 
the fluctuations. Figure~\ref{xmax_E20} clearly shows that the 
glueballino-induced EAS drastically differ from the proton-induced
showers for glueballino masses larger than 5~GeV, and hence these
showers can be distinguished even in case of low statistics. 
In case of a lighter glueballino with $M_{\tilde G}=2$~GeV, a larger
statistics is necessary to distinguish glueballino from proton, when only
$X_{\rm max}$ measurements are used. The same conclusions can be drawn
comparing the shape of the calculated profiles for individual 
$p$- and $\tilde G$-induced EAS of energy $E_0=3.2\cdot 10^{20}$ eV
with the corresponding measurements of the Fly's Eye 
collaboration~\cite{fly}, cf. Fig.~\ref{FE}. 
In doing so we choose only those showers which reach their maxima
near the measured value $X_{\max}=815\pm 50$ g/cm$^2$. Then we
average the obtained profiles and shift them to the same position of the
shower maximum, $X_{\max}=815$ g/cm$^2$. It is easy to see that 
for gluino masses larger than
5~GeV the  shape of the calculated profile strongly
disagrees with the experimental observations.
The account for the LPM effect results in only  5\% reduction of the electron
number in the shower maximum for proton-induced EAS~\cite{kalm95}
 and has an even smaller
influence on the  $\tilde G$-induced showers due to much softer
 $\pi^0$-spectrum in the glueballino interactions. 
 
\begin{figure}
\vskip1cm
\begin{picture}(11,9)
 \put (3,0) {
 \epsfig{file=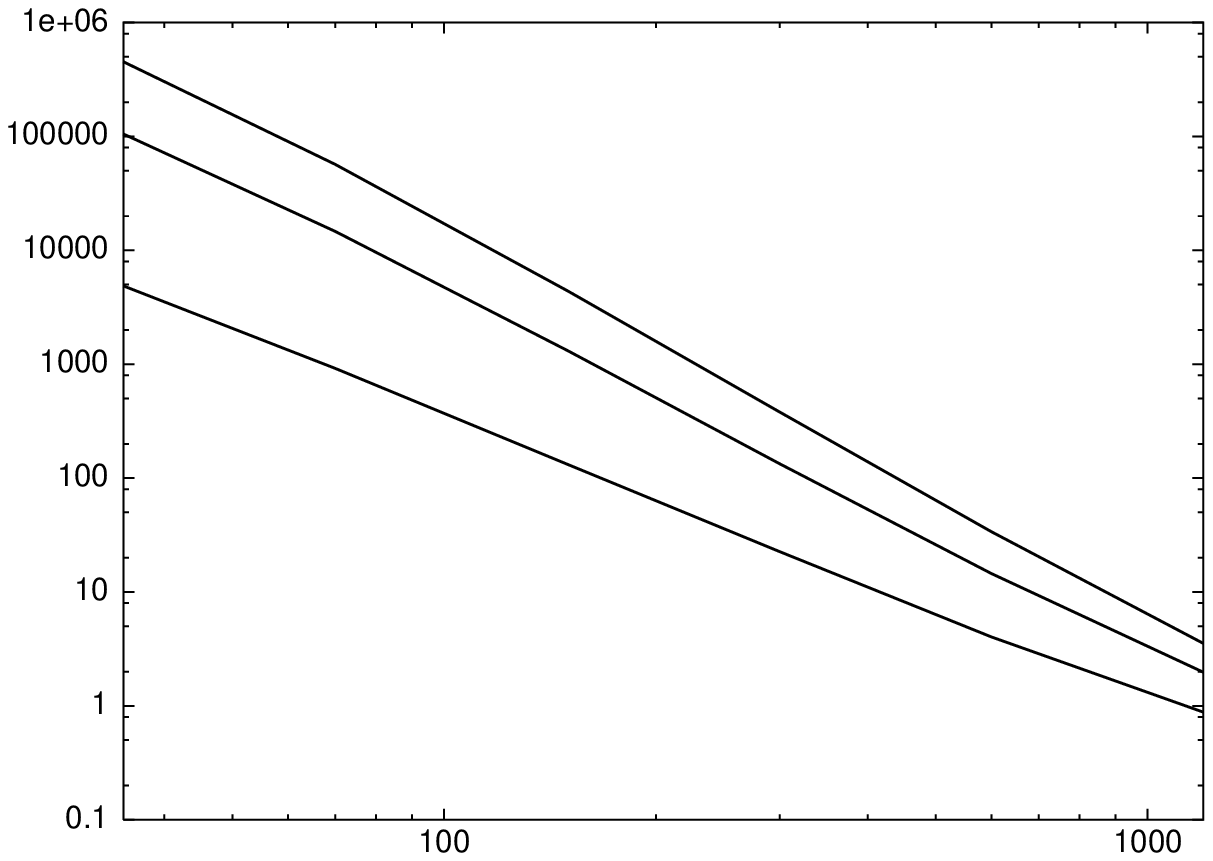,height=9.0cm,width=11.0cm}
 \put (-13.0,6.7) {proton}
 \put (-13.0,9.0) {$\tilde G$(2 GeV)}
 \put (-13.0,11.5) {$\tilde G$(5 GeV)}
 \put (-8.5,-0.6) {{\large $R$, m}}
 \put (-17.,5.0) { \begin{sideways} 
                    {\large $\rho_e(R)$, m$^{-2}$}
                    \end{sideways} }                  }
\end{picture}
\vskip0.6cm
\caption{\label{Ne}
         Electron {\it LDF} $\rho_e(R)$ in m$^{-2}$ 
         at the AKENO observation level as
         function of the distance $R$ (in m) from the shower core  
         for EAS of energy $E_0=10^{20}$~eV
         initiated by protons and by glueballinos with $M_{\tilde
         g}=2$ and 5~GeV.}
\end{figure}

\begin{figure}
\begin{picture}(11,12)
 \put (3,0) {
 \epsfig{file=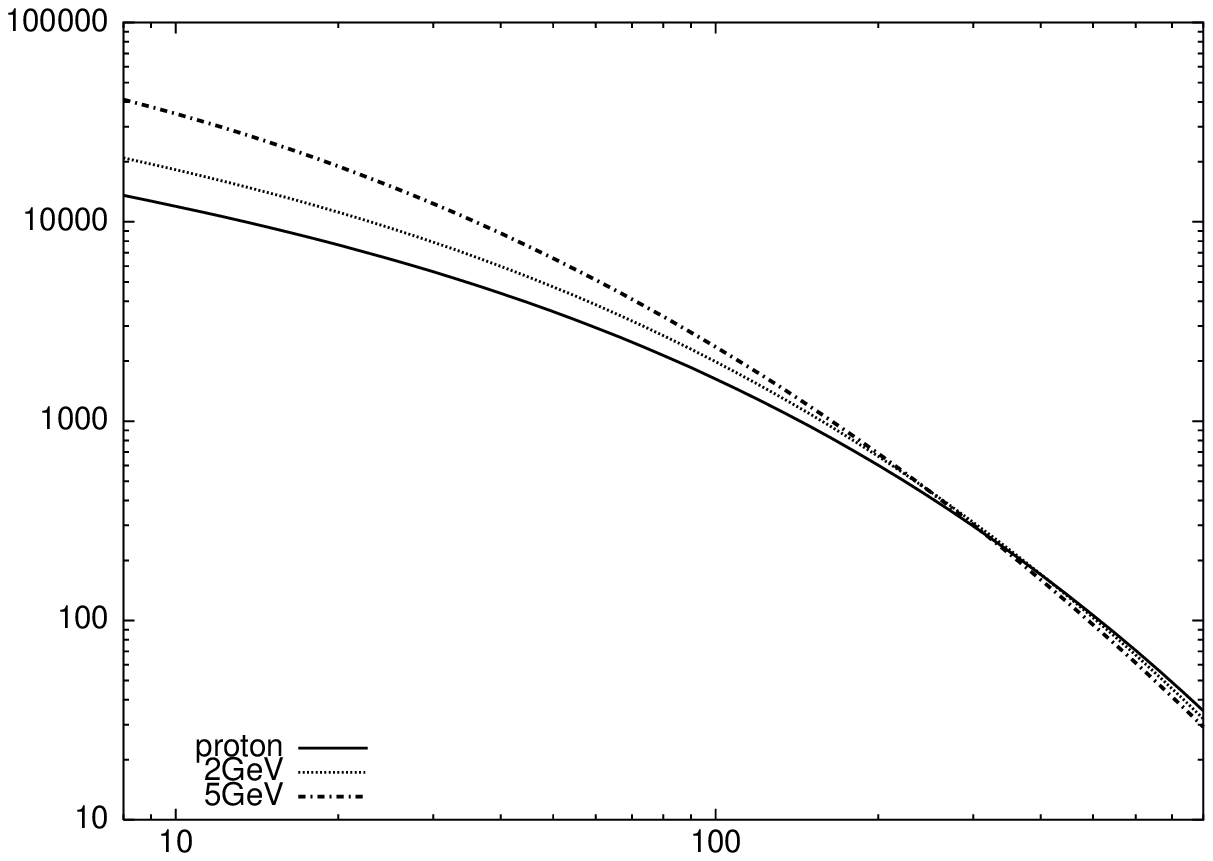,height=9.0cm,width=11.0cm}
 \put (-13.0,8.7) {proton}
 \put (-13.0,9.5) {$\tilde G$(2 GeV)}
 \put (-13.0,11.0) {$\tilde G$(5 GeV)}
 \put (-8.5,-0.6) {{\large $R$, m}}
 \put (-17.0,5.0) { \begin{sideways} 
                    {\large $\rho_{\mu}(R)$, m$^{-2}$}
                    \end{sideways} }                   }
\end{picture}
\vskip0.6cm
\caption{\label{Nmu}
         Muon {\it LDF} ($E_{\mu}>1$ GeV) $\rho_{\mu}(R)$ in m$^{-2}$ 
         at the AKENO observation level as
         function of the distance $R$ (in m) from the shower core  
         for EAS of energy $E_0=10^{20}$~eV
         initiated by protons and by glueballinos with $M_{\tilde
         g}=2$ and 5~GeV.}
\end{figure}

The calculated lateral distribution functions (LDFs) for electrons and
muons ($E_{\mu}>1$ GeV) at the AKENO observation level (900~g/cm$^2$)
are shown in Figs.~\ref{Ne} and \ref{Nmu} for the
proton and glueballinos with masses 2 and 5 GeV. The plotted values
are the LDFs of electrons and muons $\rho _{e}(R)$,
$\rho _{\mu}(R)$ at different distances $R$ from the shower core.
Although these distributions are substantially different for showers
initiated by glueballinos and protons, they hardly can be used to 
search for the light  glueballino on the basis of existing data, e.g.,
of AGASA. An adequate tool for the glueballino
search is the fluctuation of the muon density at distances $R \gsim 300$~m
from the core. This method allows in principle to discriminate 
showers initiated even by light glueballinos with $M_{\tilde G}\ap 2$~GeV
from proton-initiated EAS (see Table~\ref{fluc_mu}).

Finally, we shall compare our results with those of Albuquerque, 
Farrar and Kolb (AFK) \cite{Al98}. AFK have modified the event generator
SIBYLL including the $\tilde g$-hadron ($\tilde {G}$) as a new particle. 
The interaction properties of $\tilde g$-hadron were taken {\em ad hoc}.
Two assumptions were used for the total cross-section:
$\sigma_{\rm tot}(\tilde G p) \approx \sigma_{\rm tot}(\pi p)$  (the
favourite choice) and 
$\sigma_{\rm tot}(\tilde G p)\approx 0.1 \sigma_{\rm tot}(\pi p)$.   
The mean energy fraction transferred from the $\tilde{g}$-hadron to
the shower per interaction was modeled by a Peterson fragmentation
function. Hard interactions with the production of minijets by the
incident  $\tilde{g}$-hadron were neglected.  

It is easy to see that these modifications are not self-consistent. 
Indeed, on one hand the authors assume a large $\tilde{G}p$ cross-section,   
while on the other hand they neglect the hard processes 
(production of minijets), which give the dominant contribution to the 
$\tilde{G}p$ cross-section and make it large. In fact, our
calculations explicitly show that at ultra-high energies the
$\tilde{G}p$ cross-section and particle production are dominated by 
hard interactions for both light and heavy gluinos. For light gluinos,
valence gluon and sea partons have enough momenta for hard
interactions. For heavy gluinos, its own ("direct") hard interaction
dominates. Soft interactions without the production of parton jets are 
negligible in both cases. 

To elucidate the reason for the failure of the AFK approach, 
we have calculated the
total $\tilde G$-nucleon cross-section switching off the hard
interaction (cf. 5th entry, $\sigma_{\rm AFK}$, in Table 1 and 2):
At energies relevant for UHECR, the interactions considered by AFK
are only subdominant and result in much smaller total cross-sections
as compared with ours or those assumed by AFK.

\section{Energy losses of glueballinos on CMBR photons and glueballino
energy spectrum }

Although both valence constituents of the glueballino are electrically
neutral, UHE glueballinos loose energy due to scattering on CMBR
photons. The value of the cutoff in its energy spectrum is determined
by the transition from adiabatic energy losses (redshift) to rapidly
increasing energy losses due to the reaction 
$\tilde{G}+\gamma \to \tilde{G}+\pi^0$ at higher energies. This process 
cannot occur due to $\pi^0$ exchange in the $t$-channel. The dominant
contribution is given by the resonant formation of $\tilde{g}\bar{q}q$
states in the $s$-channel. The mass spectrum of $\tilde{g}\bar{q}q$
states was calculated as function of the gluino mass in Ref.~\cite{C}
in the MIT bag model. The lowest $\tilde g\bar qq$ state found was the 
spin-1/2 state $\tilde\rho_{1/2}$; its mass difference to the
glueballino is however, except for $m_{\tilde G}< 1.2$~GeV, too
small as to allow the decay $\tilde\rho_{1/2}\to\tilde{G}+\pi^0$,
cf.~Table~\ref{mass}. We assume therefore that the first
resonance in the $s$-channel is an excited $\tilde\rho_{1/2}^\ast$
state; for its mass $m(\tilde\rho_{1/2}^\ast)$ we use 
$m(\tilde\rho_{1/2}^\ast)=m(\tilde\rho_{1/2})+730$~MeV guided by the
mass difference between the $\rho(770)$ and the $\rho(1400)$.
The Breit-Wigner cross-section for the reaction 
$\tilde G+\gamma\to \tilde\rho_{1/2} \to \tilde G + \pi^0$ is~\cite{pdg}
\be
\sigma(s) = \frac{2\pi}{p_{\rm cm}^2} \:
            \frac{B_{\rm in}B_{\rm out}\Gamma_{\rm tot}^2 m_{\tilde\rho}^2 }
                 {(s-m_{\tilde\rho}^2)^2+(m_{\tilde\rho}\Gamma_{\rm tot})^2},
\ee
where $p_{\rm cm}$ and $\sqrt{s}$ are the momentum and the total
energy of the particles in the cm system, $m_{\tilde{\rho}}$ is the mass
of the $\tilde\rho_{1/2}$/$\tilde\rho_{1/2}^\ast$ 
particle and $\Gamma_{\rm tot}$ is its total
decay width. Finally, $B_{\rm in}$ and $B_{\rm out}$ denote the
branching ratios of the resonance to the initial and final states.

For the determination of the unknown branching ratios $B_{\rm in}$,
$B_{\rm out}$ and the total decay rate $\Gamma_{\rm tot}$, we can use
the analogy between the glueballino and the pion together with scaling
arguments as in Sec.~\ref{qgsjet} both for the $\tilde\rho_{1/2}$ and
the $\tilde\rho_{1/2}^\ast$. Then 
\be
 B_{\rm out} = \Gamma(\tilde\rho_{1/2} \to \tilde G + \pi^0)
             /  \Gamma_{\rm tot}(\tilde\rho_{1/2}) \simeq 1
\ee
and
\be 
 B_{\rm in} = \frac{\Gamma(\tilde\rho_{1/2} \to\tilde G+\gamma)}
                   {\Gamma_{\rm tot}(\tilde\rho_{1/2})}
            \simeq
              \frac{\Gamma(\rho^0 \to\pi^0+\gamma)}{\Gamma_{\rm tot}(\rho)}
            \simeq 7\cdot 10^{-4} \,.
\ee
The total decay rate $\Gamma_{\rm tot}(\tilde\rho_{1/2})\propto
g_{\tilde\rho}^2 m_{\tilde\rho}$ can be calculated using
$m_{\tilde\rho}$ from Ref.~\cite{C} together with  
\be
 g_{\tilde\rho} = \frac{m_\rho}{m_{\tilde\rho}} \: g_\rho \,.
\ee

The maximum of the photo-pion production cross-section for the
glueballino is at least a factor 8 smaller than for the proton due to
the differences in $B_{\rm in}$, the spin factors and $p_{\rm cm}$.
We neglect multipion production 
$\tilde G+\gamma\to\tilde G +\pi^+ + \pi^-$, which
operates at energies above those we are interested in. 
The non-resonant contributions have been calculated in the vector
dominance model and give only a small contribution to the total
cross-section of the glueballino. The hard processes with gluon
exchange are important only at high energies. 

The  energy loss of a particle scattering on CMBR photons
is given by~\cite{bbgdp}
\be  
  - \frac{1}{E}\:\frac{\d E}{\d t} = \frac{T}{2\pi^2\Gamma^2} \:
  \int_{E_{\rm th}}^{\infty} \d E_r \:\sigma(E_r) y(E_r) E_r
  \left\{ -\ln\left[ 1-\exp\left( -\frac{E_r}{2\Gamma T}\right) \right]
  \right\} \,.
\ee
Here, $\Gamma$ is the Lorentz factor of the particle in the CMBR frame, 
$E_r$ is the energy of a CMBR photon in the rest system of the
glueballino, $E_{\rm th}= m_{\pi}(1+m_{\pi}/2m_{\tilde{G}})$ is the 
threshold energy in the glueballino rest system and 
$y$ is the average fraction of energy lost by the glueballino in a
collision, which for one-pion production is 
\be
y(E_r)= \frac{E_r}{m_{\tilde{G}}}\frac{1+m_{\pi}^2/2E_r m_{\tilde{G}}}
{1+ 2E_r/m_{\tilde{G}}} \,.
\ee

\begin{figure}
\vskip0.6cm
\begin{picture}(11,12)
 \put (3,0) {
 \epsfig{file=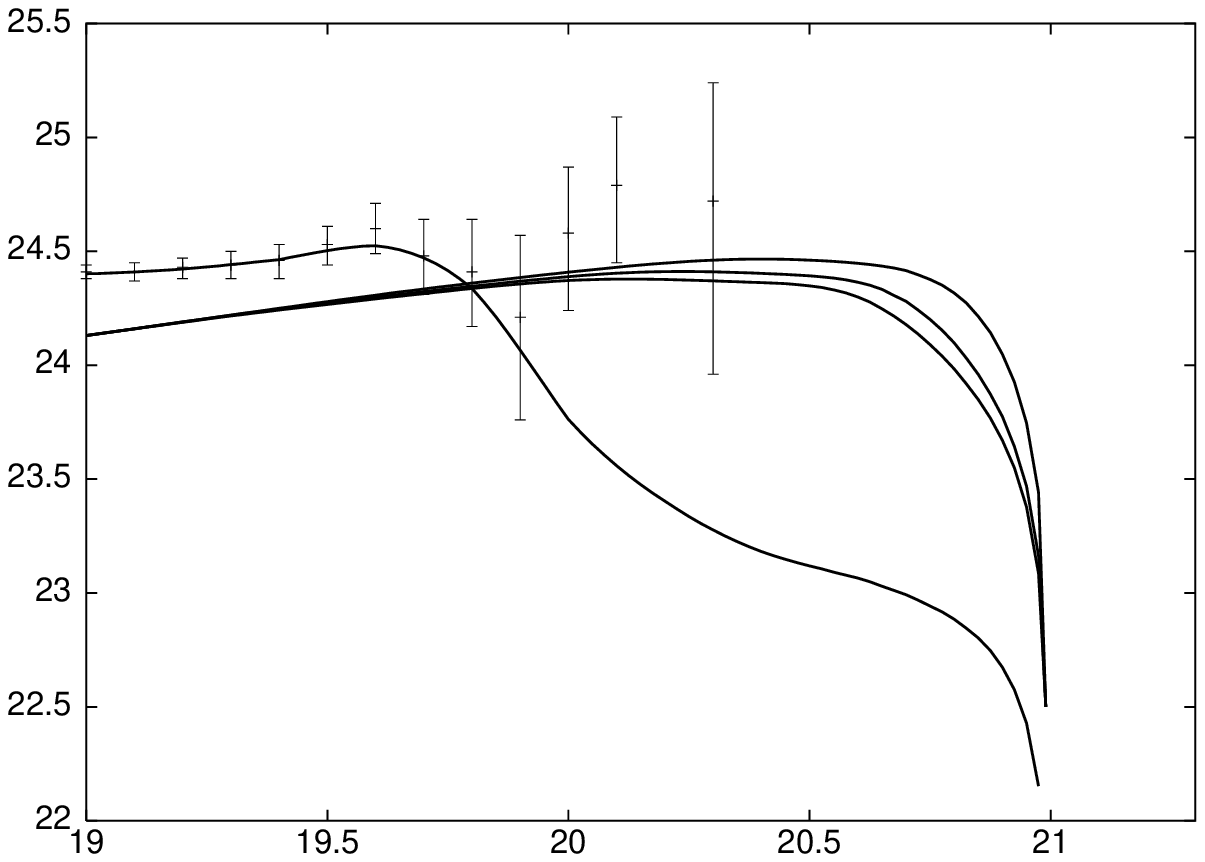,height=9.0cm,width=11.0cm}
 \put (-5.0,1.5) {proton}
 \put (-6.0,7.2) {1.5 GeV}
% \put (-3.3,7.7) {2}
 \put (-3.3,8.5) {3 GeV}
 \put (-9.1,-0.6) {$\log(E/{\rm eV})$}
 \put (-16.7,3.8) {\begin{sideways}
                   {$\log(E^3 J(E))/$m$^{-2}$s$^{-1}$eV$^{2}$}
                   \end{sideways}}   }
\end{picture}
\vskip0.4cm
\caption{\label{spec_E21}
  Diffuse glueballino flux from uniformly distributed sources with
  injection spectra $\d N/\d E\propto E^{-2.7}$ and intrinsic cutoff
  $E=10^{21}$~eV for $M_{\tilde G}=1.5$, 2, and 3~GeV, compared with
  proton flux and observational data from AGASA. Either the gluino
  flux or a combination of proton and gluino fluxes can fit the data.} 
\end{figure}

The energy spectrum $\d N/\d E$ of glueballinos emitted by diffuse
sources is at the present redshift $z=0$ given by 
%
%\be
% \frac{\d N}{\d E}(t_0) = a \int_{0}^{t_0} \d t_g \: \frac{\d N}{\d E_g}(t_g)
%                          \, \frac{\d E_g(E,t_g)}{\d E} \,,
%\ee
%
%
\be
 \frac{\d N}{\d E}(0) = a \int_{0}^{z_{\rm max}} \d z_g \:
                          (1+z_g)^{-5/2} \: 
                          \frac{\d N}{\d E_g}(z_g) \, 
                          \frac{\d E_g(E,z_g)}{\d E} \,,
\ee
where $\d N/\d E_g (t_g)$ is the injection spectrum at redshift $z_g$ and
$a$ is a constant depending mainly on the source emissivity.
We have calculated $\d E_g(E,z_g)/\d E$ as it is described in
Ref.~\cite{BeGr88}. For the injection spectrum we have used 
$\d N/\d E_g (z_g) \sim E_g^{-2.7}$ with a maximal redshift 
$z_{\rm max}=2$ and both without an intrinsic energy cutoff
(Fig.~\ref{spec_Einf}) 
and with a cutoff at $E=1\cdot 10^{22}$~eV (Fig.~\ref{spec_E21}). 

The calculated diffuse spectra are presented for three different
masses of the glueballino, $m_{\tilde G}= 1.5$, 2 and 3~GeV, 
and for comparison the corresponding proton spectra are
also shown. The cutoff in the glueballino spectra is shifted to larger
energies not only because of $m_{\tilde G}>m_N$ but also because of
the large mass gap between $\tilde G$ and
$\tilde\rho_{1/2}^\ast$. Moreover, the smaller cross-section and
energy transfer make the cutoff less pronounced.
The calculated spectra are for all three masses in agreement with 
the observations~\cite{AGASA}. However, the observed flux above the
GZK-cutoff can only be reproduced, if either the glueballino injection
spectrum is rather flat or if also at energies $E\sim 10^{19}$~eV 
glueballinos are a non-negligible component of the cosmic ray flux. 
Finally, we stress that the exact form of the glueballino energy
spectrum depend rather strongly on the mass spectrum of the hadronic
bound states of the gluino; still our general observations (shift of
the cutoff to larger energies, less pronounced cutoff) are independent
of the details of the mass spectrum.

\section{Experimental limits from beam-dump experiments}
The experiment FNAL-E-0330 by Gustafson {\it et al.} was designed to 
search for
neutral hadrons with masses $\gsim 2$~GeV and lifetimes 
$\gsim 10^{-7}$~s~\cite{gustaf}. 
In contrast to the recent experiments~\cite{Ad97,Fa99,Al99,Al97},
the Gustafson experiment was sensitive to long-lived or stable hadrons.
It set upper limits on the production cross-section of
these hadrons in the mass range 2--12~GeV which are given in the second
line of Table~\ref{gustaf}. However, the relation between the measured
mass, $M_{\rm meas}$, in the Gustafson experiment and the physical
mass of $\tilde{g}$-hadrons needs careful consideration. 

In the Gustafson experiment, short pulses of protons with energy 300~GeV
produced in the beam-dump target neutral secondary particles, which were
detected in a calorimeter at distance $l=590$~m from the target. 
The calorimeter had the total thickness $X_{\rm cal}= 900$~g/cm$^2$,
and it was assumed that a particle looses all its energy therein. Then
the measured difference in the time of flight $\delta t$ of massive
secondaries and photons determines the mass of the secondary,
\be
M_{\rm meas}= (2c E_{\rm loss}^2\delta t/l)^{1/2} \,.
\label{mmeas}
\ee
Here, the energy loss $E_{\rm loss}$ in the calorimeter is taken as the
energy of a particle. In fact, a $\tilde{g}$-hadron looses however only
a part of its energy in the calorimeter: 
$E_{\rm loss}/E=K_{\rm inel}X_{\rm cal}/X_{\rm int}$, where $X_{\rm int}$
is the inelastic interaction length. Then the physical mass $M_{\tilde G}$ 
of a $\tilde{g}$-hadron is connected with the measured mass as
\be
M_{\rm meas}=(E_{\rm loss}/E)M_{\tilde G} \,.
\label{ratio}
\ee
In the case of a glueballino, one obtains from Eq.~(\ref{ratio})
together with the data of Table~\ref{tab1} $M_{\rm meas} \sim 1$~GeV 
for $M_{\tilde G}=2$ and 5~GeV. Thus,
in both cases glueballinos fall in the region of neutron 
background and therefore these masses are allowed. 

Are heavier masses of  $\tilde{g}$-hadrons allowed by the Gustafson
experiment? 

We have calculated the cross-section for gluino production in 
pp-collision via the two subprocesses $gg\to\tilde g\tilde g$ and 
$q\bar{q} \to\tilde g\tilde g$ at next-to-leading order~\cite{Be97} 
for $E_{\rm lab}=300$~ GeV using the program {\tt Prospero}~\cite{pro}.  
As parton distributions we have chosen those of
Gl\"uck, Reya and Vogt~\cite{GRV}, while we have used $Q=m_{\tilde g}$
as renormalization and factorization scales.
In Table~\ref{gustaf}, the production cross-sections averaged over the
region allowed by the experimental cuts are compared with the upper
limits from the Gustafson experiment. The
inspection of Table ~\ref{gustaf} for larger masses shows that the 
calculated cross-sections are smaller than the upper limits given in 
the second line. 
We conclude that the results of the Gustafson experiment do not
exclude $\tilde{g}$-hadrons.

A modified Gustafson experiment has great potential to discover
light  $\tilde{g}$-hadrons in the window 2--4~GeV or to reliably
exclude it. For this aim, a thicker calorimeter is needed. If
$\tilde{g}$-hadrons lose all their energy in the calorimeter, then 
$M_{\rm meas}\approx M_{\tilde G}$ and the third and fourth line
of Table~\ref{gustaf} show that the calculated cross-sections for 
$M_{\tilde G}$ in the interval 2--4~GeV are larger than the experimental
upper bounds. It means that $\tilde{g}$-hadron in this mass window
can be discovered or excluded, if the calorimeter is thick enough. 
Taking into account the large penetrating power of 
$\tilde{g}$-hadrons the neutron background can be greatly reduced
by  placing the absorber behind the target which thickness is tuned to 
absorb the neutrons but to be transparent for $\tilde{g}$-hadrons.  

The signature of $\tilde{g}$-hadrons is given by the compatibility of
the gluino production cross-section with the measured flux of detected
particles.  The path-length and the average fraction of lost energy 
will serve as further indicators. The measured properties of
$\tilde{g}$-hadrons will then allow reliable calculations of
$\tilde{g}$-hadron production in astrophysical sources and their
detection in EAS.

\section{Conclusions}
There are two mass windows where a light gluino is marginally  allowed: 
$m_{\tilde{g}} \lsim 3$~GeV and  $25\lsim m_{\tilde g} \lsim 35$~GeV.
The first window is disfavoured by the gluino contribution to 
the running of $\alpha_s$ and to the colour coefficients. The 
lightest $\tilde{g}$-hadron is heavier than the gluino by the mass of 
constituent gluon or quarks. The existence of a
light {\em quasi-stable\/} $\tilde{g}$-hadron, corresponding 
to the first window for gluino masses, crucially depends on 
whether the neutral or charged  $\tilde{g}$-hadron is the lightest
one. A light quasi-stable charged $\tilde{g}$-hadron is forbidden by
the production of ``wild hydrogen'' in the Earth atmosphere
by cosmic rays. The case that the lightest $\tilde{g}$-hadron is neutral is
also forbidden, if it forms a bound state with the proton (``wild deuteron''). 
The status of the light gluino and lightest $\tilde{g}$-hadron can further
be clarified by further theoretical analysis. At present we
conservatively consider the first gluino window as disfavoured, but
not excluded.  

As the carrier of UHE signal, the light $\tilde{g}$-hadrons 
with a mass larger  than 1.5~GeV have a spectrum with the GZK cutoff 
beyond the observed energy range (see Fig.~\ref{spec_Einf}). Together with
accelerator limits on the gluino mass, $m_{\tilde g}< 3$~GeV,
it leaves for $\tilde{g}$-hadron masses the
narrow window 1.5--4~GeV. The other window allowed by accelerator
experiments is 25--35~GeV. 

In this paper we have studied the interaction of $\tilde{g}$-hadrons
with nucleons and nuclei and the development of UHE EAS initiated by such
particles. In practice, we have considered the special case of a
glueballino as the primary particle. We think, however,
that any other  $\tilde{g}$-hadron with equal mass has essentially 
the same interaction properties. We have
calculated the glueballino-nucleon inelastic and total cross-sections 
for different masses of glueballinos (see Figs.~\ref{tot}, \ref{inel}). 
The longitudinal development of the EAS with energy $3\cdot 10^{20}$~eV 
in the atmosphere is shown in Fig.~\ref{FE}. One can
see that glueballinos heavier than 5~GeV resemble penetrating particles. 
The profiles shown in the Figure are directly measured in Fly's Eye
experiment, and the data are quite different from the graphs displayed
here for heavy glueballinos. The second mass window 25--35~GeV can
be already reliably excluded on
the basis of these measurements, though a detailed analysis is desirable. 

The showers produced by  $\tilde{g}$-hadron from the low-mass window 
are similar to proton-initiated showers in all characteristics (see
the longitudinal profiles in Figs.~\ref{pro_E20}, \ref{pro_E19} and
\ref{pro_E17}, the distributions over $X_{\max}$
in Figs.~\ref{xmax_E20}, \ref{xmax_E19} and \ref{xmax_E17}, 
lateral distributions for electrons and muons in Figs.~\ref{Ne}, \ref{Nmu} 
and fluctuations in muon and electron densities in Tables~\ref{fluc_e} 
and \ref{fluc_mu}). In principle, the best possibility to distinguish the
showers produced by  $\tilde{g}$-hadron with mass 2~GeV from the 
proton-induced showers are the fluctuations in the muon density 
at large distances $d> 600$~m (see Table~\ref{fluc_e}). However, the
statistics of the largest array at present, AGASA, is not sufficient for
such a discrimination. 
The large $X_{\max}$ tail in the distribution of showers initiated by
$\tilde{g}$-hadrons over $X_{\max}$ offers for HiRes or AUGER another,
more promising possibility to (dis-) prove light gluinos as UHE
primaries.

More reliably the quasi-stable 2--3~GeV  $\tilde{g}$-hadron can be found in 
an especially designed accelerator beam-dump experiment (see section 5).
In this experiment, it is possible to discover supersymmetry
(light gluino) and to find a new primary for the UHE cosmic signal.

\section*{Acknowledgements}
We are grateful to Michael Spira for helpful comments about the use of
{\tt PROSPINO}.

%%%%%%%%%%%%%%%%%%%%%%%%%%%%%%%%%%%%%%%%%%%%%%%%%%%%%%%%%%%%%%%%%%%%%%%%%%%%

\newpage

\begin{table}
\begin{tabular}{c||c||c|c|c|c}
   & $\pi$ & $M_{\tilde G}=2$~GeV & $M_{\tilde G}=5$~GeV
& $M_{\tilde G}=10$~GeV & $M_{\tilde G}=50$~GeV
\\[0.5ex] \hline  
$\sigma_{\rm tot}$ & 24.4 & 3.9 & 2.9 & 2.4 & 2.2 \\
$\sigma_{\rm in}$ & 20.9 & 3.8 & 2.8 & 2.4 & 2.1 \\
$\sigma_{\rm tot}^{\rm s-coupl}$ & & 3.8  & 2.7 &  2.4 & 2.2 \\
$\sigma_{\rm tot}^{\rm d-coupl}$ & & 0.14 & 0.15 & 0.013 & 0 \\ 
$\sigma_{\rm AFK}$ &   & 3.8 & 2.7 & 2.4 & 2.2 \\
$K_{\rm inel}$ & & 0.25 & 0.15 & 0.074 & 0.0054  
\end{tabular}
\caption{\label{tab1} 
Total cross-section $\sigma_{\rm tot}$, 
inelastic cross-section $\sigma_{\rm in}$, 
cross-section due to soft coupling $\sigma_{\rm tot}^{\rm s-coupl}$, 
cross-section due to direct coupling $\sigma_{\rm tot}^{\rm d-coupl}$, 
cross-section without hard interactions $\sigma_{\rm AFK}$ and the
inelasticity coeffiecent $K_{\rm inel}$ for the glueballino-nucleon
scattering. The data are given for four values of the glueballino mass
$M_{\tilde G}$ from 2~GeV to 50~GeV at glueballino energy 
$E_{\rm lab}=100$~GeV. As references, the total and inelastic
cross-section of the pion are also given. All cross-sections are in
mbarn.} 
\end{table}

\begin{table}
\vskip2.3cm
\begin{tabular}{c||c||c|c|c|c}
   & $\pi$ & $M_{\tilde G}=2$~GeV & $M_{\tilde G}=5$~GeV
& $M_{\tilde G}=10$~GeV & $M_{\tilde G}=50$~GeV
\\[0.5ex] \hline  
$\sigma_{\rm tot}$ & 152 & 103 & 94.9 & 91.3 & 88.0 \\
$\sigma_{\rm in}$ & 112 & 71.3 & 65.9 & 63.4 &  60.2 \\
$\sigma_{\rm tot}^{\rm s-coupl}$ & & 85.5 & 69.4 &  61.8 & 50.8 \\
$\sigma_{\rm tot}^{\rm d-coupl}$ & & 72.1 & 72.4 & 72.3 & 74.4 \\  
$\sigma_{\rm AKF}$ & 92.2 & 18.2 & 13.3 & 11.8 & 10.6 \\
$K_{\rm inel}$ & & 0.26 & 0.14 & 0.082 & 0.018 
\end{tabular}
\caption{\label{tab2} The same as Table~1 for $E_{\rm lab}=10^{12}$~GeV.}
\end{table}

\begin{table}
\vskip2.3cm
\begin{tabular}{l||c|c|c|c|c|c|c}
\hspace{5mm} R, m & 10 & 50 & 100 &200 & 300~m & 600~m & 1200~m  
\\[0.5ex] \hline  
proton &             0.077 & 0.054 & 0.068 & 0.095 & 0.11 & 0.15 & 0.18 \\
$\tilde G$(2~GeV)  &   0.20 & 0.25 & 0.27 & 0.30 & 0.31 & 0.34 & 0.36 \\ 
$\tilde G$(5~GeV)  &   0.33 & 0.37 & 0.39 & 0.41 & 0.43 & 0.45 & 0.47 \\
$\tilde G$(10~GeV) &   0.40 & 0.43 & 0.45 & 0.47 & 0.48 & 0.50 & 0.52  
\end{tabular}
\caption{\label{fluc_e} Normalized $1\sigma$ fluctuations of the electron 
LDF $\sigma_{\rho_{e}(R)}/\rho_{e}(R)$ 
at the AKENO observation level for EAS of energy $E_0=10^{20}$~eV
initiated by  protons and 
glueballinos of  masses 2, 5 and 10~GeV.}
\end{table}

%%%%%%%%%%%%%%%%%%%%%

\begin{table}
\begin{tabular}{c||c|c|c|c|c|c}
\hspace{5mm} R, m & 10 &  100 &200 & 300~m & 600~m & 1200~m  
\\[0.5ex] \hline  
proton &              0.19 & 0.16 & 0.16 & 0.17 & 0.20 & 0.25  \\
 $\tilde G$(2~GeV)  &   0.18 & 0.20 & 0.24 & 0.27 & 0.33 & 0.41 \\
$\tilde G$(5~GeV)   &   0.23 & 0.30 & 0.34 & 0.37 & 0.43 & 0.51 \\
$\tilde G$(10~GeV)  &   0.30 & 0.36 & 0.40 & 0.42 & 0.48 & 0.55 \\
\end{tabular}
\caption{\label{fluc_mu} Normalized $1\sigma$ fluctuations of the muon 
LDF ($E_{\mu}>1$~GeV)  $\sigma_{\rho _{\mu}(R)}/\rho _{\mu}(R)$ 
at the AKENO observation level for EAS of energy $E_0=10^{20}$~eV
initiated by  protons and 
glueballinos of  masses 2, 5 and 10~GeV.}
\end{table}

\begin{table}
\vskip2.3cm
\begin{tabular}{c||c|c|c|c|c|c}
$M_{\rm meas}/$GeV & 2 & 3 & 4 & 6 & 8 & 10   \\
\hline  
90\% C.L. & 1.0e-32 & 3.6e-33 & 1.3e-33& 5.3e-33& 2.0e-33 & 1.7e-33 \\
\hline   \hline   
$m_{\tilde g}$/GeV & 2 & 3 & 4 & 6 & 8 & 10   \\
\hline
$E\d^3\sigma/\d^3p$  & 3.8e-30 & 1.7e-31 & 1.1e-32& 4.2e-35& 3.9e-38
& 8.9e-43  
\end{tabular}
\caption{\label{gustaf} The upper 90\% C.L. limit on the
  invariant cross-sections $E\d^3\sigma/\d^3p$ (second line) for
  measured masses $M_{\rm meas}$ (first line) given by the Gustafson
  experiment. The fourth line gives the calculated invariant
  cross-sections for gluino production in pp-interaction 
  at $E_{\rm p,lab}=300$~GeV and for gluino mass $m_{\tilde g}$ (third
  line).  All cross-sections are in cm$^2$/GeV$^2$.}
\end{table}

\begin{table}
\vskip2.3cm
\begin{tabular}{c||c|c|c|c}
$m_{\tilde g}/$GeV & 0.64 & 1.04 & 1.48 & 2.41  \\ \hline
$M_{\tilde G}/$GeV & 1.00 & 1.50 & 2.00 & 3.00  \\ \hline
$m_{\tilde\rho_{1/2}}/$GeV & 1.17 & 1.58 & 2.04 & 3.03 
\end{tabular}
\caption{\label{mass} The masses of the glueballino $\tilde G$ and
  the $\tilde\rho_{1/2}$ as function of the gluino mass $m_{\tilde g}$
  according to Ref.~\cite{C}.}
\end{table}

\clearpage
\newpage
\unitlength0.7cm

%%%%%%%%%%%%%%%%%%%%%

%%%%%%%%%%%%%%%%%%%%%

%%%%%%%%%%%%%%%%%%%%%%%%%%%%%%%

%%%%%%%%%%%%%%%%%%%%%%%%%%

%%%%%%%%%%%%%%%%%%%%%%%%%%

%%%%%%%%%%%%%

\end{document}